\documentclass[useAMS,usenatbib]{mn2e}
\usepackage{epsfig}

\newcommand{\ci}{[C\,{\sc i}]}

\title{Is atomic carbon a good tracer of molecular gas in metal-poor galaxies?}
\author[Glover \& Clark]
{Simon C. O. Glover$^{1,}$\thanks{E-mail: glover@uni-heidelberg.de} \& Paul C. Clark$^{2}$ \\
$^{1}$ Universit\"at Heidelberg, Zentrum f\"ur Astronomie, Institut f\"ur Theoretische Astrophysik, \\ Albert-Ueberle-Strasse 2,  69120 Heidelberg, Germany \\
$^{2}$ School of Physics and Astronomy, Queen's Buildings, The Parade, Cardiff University, Cardiff, CF24 3AA \\
}

\begin{document}
\maketitle

\begin{abstract}
  Carbon monoxide (CO) is widely used as a tracer of molecular hydrogen (H$_{2}$) 
   in metal-rich galaxies, but is known to become ineffective in low metallicity
   dwarf galaxies.  Atomic carbon has been suggested as a superior tracer of H$_{2}$ 
   in these metal-poor systems,  but its suitability remains unproven.  To help us to 
   assess how well atomic carbon traces H$_{2}$ at low metallicity, we have performed
   a series of numerical simulations of turbulent molecular clouds that cover a 
   wide range of different metallicities.  Our simulations demonstrate that in star-forming clouds, 
   the conversion factor between [CI] emission and H$_{2}$ mass, $X_{\rm CI}$, 
   scales approximately as $X_{\rm CI} \propto {\rm Z}^{-1}$. We recover
   a similar scaling for the CO-to-H$_{2}$ conversion factor, $X_{\rm CO}$, but find 
   that at this point in the evolution of the clouds, $X_{\rm CO}$ is consistently smaller
   than $X_{\rm CI}$, by a factor of a few or more. 
   We have also examined how $X_{\rm CI}$ and $X_{\rm CO}$ evolve with time.
   We find that $X_{\rm CI}$ does not vary strongly with time, demonstrating
   that atomic carbon remains a good tracer of H$_{2}$ in metal-poor systems
   even at times significantly before the onset of star formation. On the other hand,
   $X_{\rm CO}$ varies very strongly with time in metal-poor clouds, showing that
   CO does not trace H$_{2}$ well in starless clouds at low metallicity.
\end{abstract}

\begin{keywords}
galaxies: ISM -- ISM: clouds -- ISM: molecules --  stars: formation
\end{keywords}

\section{Introduction}
Giant molecular clouds (GMCs) play a central role in the life of our own and other galaxies,
as they are the sites where essentially all star formation occurs. For this reason, the study of 
the formation and evolution of GMCs is a highly important topic. However, this study is hampered 
by the fact that the two main constituents of the gas making up a  GMC -- 
molecular hydrogen (H$_{2}$) and atomic helium -- do not emit radiation at the low gas
temperatures present in most GMCs. In order to study the molecular gas, it is therefore necessary
to observe emission from less abundant chemical species that do emit at low temperatures and
that are correlated with the H$_{2}$ distribution. 

One of the most popular such tracer species is carbon monoxide (CO). This has several advantages
as a tracer of molecular gas. First, there is a lot of it in most GMCs: it is typically the second most
abundant molecular species, after H$_{2}$. Second, it emits readily at low temperatures: the 
energy separation between the rotational ground state ($J = 0$) and the first rotationally excited level
($J = 1$) is only 5.5~K. In addition, the main chemical pathways that lead to the formation of
CO in the interstellar medium all involve H$_{2}$ as a key ingredient at some point  \citep[see e.g.\
the detailed discussion of CO formation chemistry in][]{glo10}, and so we would not expect large
amounts of CO to form in regions devoid of H$_{2}$.

Observationally, CO emission has indeed proved to be a good tracer of molecular 
gas within the Milky Way. A number of different studies, summarized in the recent review
by \citet{bwl13}, have shown that on scales comparable to or larger than individual GMCs there is
a good correlation between the velocity-integrated intensity in the $J = 1 \rightarrow 0$
line of $^{12}$CO and the H$_{2}$ column density. 	The conversion factor from integrated
intensity to column density -- the so-called X-factor -- is given approximately by
\begin{equation}
X_{\rm CO, MW} = 2 \times 10^{20} \: {\rm cm^{-2} (K \, km/s)^{-1} }  \label{xcomw}
\end{equation}
for Milky Way GMCs, and similar values are also found for other nearby, Milky Way-like
disk galaxies. Simulations of GMCs in environments similar to that of the local ISM
typically yield values for $X_{\rm CO}$ close to this observationally determined value
\citep[see e.g.][]{gm11,shetty11a,shetty11b,nara11,nh13}. Its lack of
significant variation from cloud to cloud can be understood as a consequence of the
fact that the main cloud properties that determine $X_{\rm CO}$ in metal-rich galaxies 
-- the velocity dispersion, the temperature and the surface density of the gas -- do not
vary much from cloud to cloud \citep{nh13}.

However, in low metallicity systems such as dwarf galaxies, CO is much less useful
as a tracer of molecular gas. A growing number of observational studies have provided
evidence that $X_{\rm CO}$ increases substantially as the metallicity decreases 
\citep[see e.g.][]{israel97,madden97,leroy07,leroy11,gratier10,bolatto11,smith12},
owing to the reduction in the mean extinction of the molecular gas \citep[see e.g.][]{gm11}.
Numerical simulations also suggest that the $X_{\rm CO}$ may become strongly 
time-dependent at low metallicity (\citealt{gc12b}; see also Section~\ref{sec:time_evol}
below). Furthermore, CO has proved to be 
extremely difficult to detect in very low metallicity systems: at ${\rm Z} \leq 0.1 \: {\rm Z_{\odot}}$,
there are secure detections of CO in only two systems, the Wolf-Lundmark-Melotte
and Sextans A dwarf galaxies \citep{elm13,shi15}.

In view of the difficulties involved with using CO as a tracer of molecular 
gas in metal-poor galaxies, it is natural to ask whether there are better tracers 
that we might use. One interesting possibility is neutral atomic carbon, C. This 
produces two fine structure lines in its electronic ground state, the 
$^{3}P_{1} \rightarrow \mbox{}^{3}P_{0}$ transition at 609$\mu$m and the 
$^{3}P_{2} \rightarrow \mbox{}^{3}P_{1}$ transition at 370$\mu$m. Its potential as 
a tracer of molecular gas was championed by \citet{ptv04}, who pointed out that 
although simple PDR models predict that it will be confined to a thin transition 
region at the edge of molecular clouds \citep[see e.g.][]{langer76},
observations do not bear this out, and have long shown that C and CO
emission are surprisingly well correlated in local clouds 
\citep{frerking89,little94,sch95,kra08}. This behaviour can be understood as a
consequence of clumpy substructure within molecular clouds: CO emission 
comes primarily from within clumps and filaments, while \ci~emission traces 
their surfaces, but since the typical scale of the structures is small and there
are often many along each line of sight, the result is that the projected CO
and \ci~emission are well-correlated, both spatially and also in terms of
velocity \citep{stut88,genz88,pjk94,kram04,glo15}. 

More recently,  \citet{offner14} have explored the usefulness of atomic
carbon as a molecular gas tracer in solar metallicity clouds. They used
the {\sc 3d-pdr} code \citep{bisbas12} to post-process simulations of turbulent
molecular clouds, allowing them to predict the chemical composition and
temperature of the gas at each point in the cloud. Based on this,
they then computed synthetic emission maps for the 609$\mu$m of
\ci~and the $J = 1 \rightarrow 0$ transition of CO, using the {\sc radmc-3d} radiation transfer 
code.\footnote{http://www.ita.uni-heidelberg.de/$\sim$dullemond/software/radmc-3d/}
They found that \ci~emission was indeed a good tracer of H$_{2}$ column density
in their model clouds. They also computed a mean \ci-to-H$_{2}$ conversion factor,
$X_{\rm CI}$, for their models, and showed that $X_{\rm CI} = 1.1 \times 10^{21}
\: {\rm cm^{-2} (K \, km/s)^{-1} }$ for a cloud with solar metallicity illuminated by
the standard interstellar radiation field (ISRF), and $X_{\rm CI} = 1.2 \times 10^{21}
\: {\rm cm^{-2} (K \, km/s)^{-1} }$ for a solar metallicity cloud illuminated by a ten
times stronger ISRF. However, \citet{offner14} did not explore whether $X_{\rm CI}$
is sensitive to the metallicity of the gas. In addition, there are a couple of
inconsistencies in the \citet{offner14} approach that may influence their results.
Firstly, they perform their hydrodynamical simulations assuming an isothermal
equation of state \citep{offner13}, and only solve for the actual temperature distribution
during the later post-processing step. This is a reasonable approximation in dense,
well-shielded gas, but not in the lower density, moderately shielded gas where
atomic carbon dominates (c.f.\ their Figures 6 and 7). Secondly, they evolve the chemistry
for a period of 10~Myr, but stop the hydrodynamical simulation after $\sim 1 \: {\rm Myr}$.
They may therefore miss important non-equilibrium effects in the chemistry.

In \citeauthor{glo15}~(2015; hereafter, Paper I), we presented results from
a study of the \ci~emission produced by a model of a turbulent molecular cloud
that does not suffer from these inconsistencies. In our simulation, we evolved
the chemistry and internal energy of the gas simultaneously with the hydrodynamics,
allowing us to capture any influence of the non-isothermal equation of state on 
the dynamical evolution and to ensure that the chemistry is not evolved for a longer
period of time than the dynamics. The cloud that we modeled in Paper I had a solar metallicity, and was illuminated by the standard
\citet{dr78} ultraviolet (UV) radiation field. We found, in agreement with \citet{offner14}, that
the \ci~emission line was a good tracer of the molecular mass of the cloud. 


In the present paper, we extend this study by examining clouds with a range of different 
metallicities. For each metallicity, we run models with several different values for the
strength of the ISRF. Our goal is to understand how well \ci~emission can be used as a 
tracer of molecular gas in this wider range of systems, and whether it becomes clearly 
superior to CO at low metallicities. Since clouds in extragalactic systems are typically
unresolved or only marginally resolved, our main focus in this study is on how 
$X_{\rm CI}$ varies as a function of metallicity, radiation field strength and time.

The structure of our paper is as follows. In Section~\ref{method}, we outline the
numerical method used to perform our simulations and to post-process the results.
We also give details of the initial conditions that we used for our simulations. 
In Section~\ref{result}, we present and discuss our results. Finally, we conclude
in Section~\ref{conc}.

\section{Method}
\label{method}
\subsection{Hydrodynamical model}
To model the chemical, thermal and dynamical evolution of the gas in our model
clouds, we use a modified version of the Gadget 2 SPH code \citep{springel05}.
Our modified version of the code includes a simplified chemical model that 
follows the abundances of H, H$_{2}$, H$^{+}$, C, C$^{+}$, CO, HCO$^{+}$,
He, He$^{+}$, and free electrons
\citep{gm07a,gm07b,nl99}, a detailed cooling function, and a
treatment of the effects of attenuation of the interstellar radiation field (ISRF) based
on the {\sc treecol} algorithm \citep{cgk12}. To model gravitationally collapsing regions
that become too small to resolve, we use a sink particle algorithm based on \citet{bbp95}
and \citet{jap05}. The version of the code used here is the same as that used in 
Paper I, and a more extensive discussion of its capabilities can be found in that paper.

Although our use of a simplified chemical network introduces some uncertainty into
our results that would not be present if we used a more extensive network, the impact 
of this on our predicted C and CO abundances is likely to be small. We showed in an
earlier paper \citep{gc12c} that the simplified chemical network that we use in this
study produces comparable results for the C and CO distributions to the larger
\citet{glo10} network. That network was in turn benchmarked against  an extensive
model derived from the UMIST astrochemical database \citep{umist07} and shown to produce
results in good agreement with the latter for a range of different physical conditions.
In addition, the fact that our results agree well with those of \citet{offner14}, who
make use of a chemical network containing over 200 species and several thousand
reactions, also gives us confidence that our use of a simplified network is not a major
source of error.

\subsection{Post-processing}
We construct synthetic emission maps of our simulated clouds in the 
$^{3}P_{1} \rightarrow \mbox{}^{3}P_{0}$ fine structure line of 
\ci~(hereafter simply referred to as the \ci~$1 \rightarrow 0$ line)
and the
$J = 1 \rightarrow 0$ line of $^{12}$CO using the {\sc radmc-3d} radiation transfer 
code.
The region modelled with {\sc radmc-3d} is a cubical sub-volume of the cloud, with a 
side length of 16.2~pc that contains almost all of the CO and neutral atomic carbon
present in the model clouds. Before feeding our data to {\sc radmc-3d}, we first interpolate
it onto a Cartesian grid with a resolution of $256^{3}$ zones. In Paper I we explored the
effect of varying this resolution and showed that $256^{3}$ zones was sufficient to fully
resolve the \ci~emission.

To compute the level populations required by the code, we used the large velocity gradient 
(LVG) approximation. The implementation of this approximation within {\sc radmc-3d} is
described in detail in  \citet{shetty11a,shetty11b}. We fixed the velocity resolution of the
code at $\Delta v = 0.094 \: {\rm km \: s^{-1}}$ and accounted for small-scale unresolved
velocity fluctuations by including a microturbulent contribution to the linewidth with size
$v_{\rm mtrb} = 0.2 \: {\rm km \: s^{-1}}$. However, the values that we derived for the 
\ci~and CO velocity-integrated intensities are relatively insensitive to these choices.

\subsection{Initial conditions}
Our default initial conditions in this study are similar to those used in paper I and
in several other previous papers \citep[see e.g.][]{gc12a,gc12b}. The cloud mass
is $10^{4} \: {\rm M_{\odot}}$, and the cloud is simulated using two million SPH particles.
The resulting mass resolution is $0.5 \: {\rm M_{\odot}}$. In paper I, we explored the
effects of increasing this resolution, but found that this resulted in only minor differences
in the \ci~emission produced by the cloud. The cloud was initially spherical, with a
radius of 6.3~pc. The initial H nuclei number density of the gas was 
$n_{0} = 276 \: {\rm cm^{-3}}$, within a factor of two of the median value determined for
local GMCs by \citet{RomanDuval2010}. The initial gas temperature was $T_{\rm gas}
= 20 \: {\rm K}$ and the initial dust temperature was $T_{\rm dust} = 10 \: {\rm K}$.\footnote{Note 
that since $T_{\rm gas}$ and $T_{\rm dust}$ rapidly adjust to close to their thermal
equilibrium values following the beginning of the simulations, our results do not depend
on these starting values.}

The velocity field of the gas in the clouds was constructed as a Gaussian
random field, with a power spectrum $P(k) \propto k^{-4}$, where $k$ is the 
wavenumber. The amplitude of the velocity fluctuations was constrained to
yield an initial RMS velocity of $v_{\rm rms} = 2.8 \: {\rm km \: s^{-1}}$. This
value was chosen so that the initial kinetic energy of the cloud was equal to
its gravitational binding energy. The velocity field was then allowed to evolve
during the simulation without any further kinetic energy input. The cloud was
simulated using vacuum boundary conditions, but the artificial ``evaporation''
of material from the edge of the cloud was prevented by the application of an
external pressure term \citep{benz90}. 

We performed a series of simulations using these initial conditions, and varying
the metallicity of the gas and the strength of the imposed radiation field. The values
used in the different simulations are summarized in Table~\ref{tab:sims}. In our solar metallicity
simulations, the total abundances of carbon and oxygen relative to hydrogen were
taken to be $x_{\rm C} = 1.4 \times 10^{-4}$ and $x_{\rm O} = 3.2 \times 10^{-4}$, 
respectively \citep{sem00}. In the lower metallicity simulations, these values were
scaled by a factor ${\rm Z / Z_{\odot}}$. In most of our simulations, we also assumed
that the dust-to-gas ratio ${\cal D}$ scaled as ${\rm Z / Z_{\odot}}$. However, there is
some observational evidence that it actually falls off more rapidly as the metallicity
decreases \citep[e.g.][]{gala11,hc12,rr14}, and so we also performed three simulations 
(Z05-G1-D0375, Z02-G1-D01 and Z01-G1-D001) in which we adopted smaller dust-to-gas 
ratios. For simplicity, we neglect any effects arising due to changes in the size distribution
or composition of the dust as we change the metallicity and dust-to-gas ratio.
In all of our simulations, the oxygen and hydrogen were assumed to begin in neutral 
atomic form, and the carbon was assumed to start as C$^{+}$. 

The spectral shape of our adopted interstellar radiation field follows \citet{dr78}
in the ultraviolet and \citet{mmp83} at longer wavelengths. We consider several
different radiation field strengths, quantified by the parameter $G_{0}$ which is
normalized so that $G_{0} = 1$ corresponds to the \citet{dr78} radiation field.
In our {\sc  treecol}-based treatment of the attenuation of the radiation field,
we used a total of 48 pixels to represent the angular variation of the radiation field
around each SPH particle. In paper I, we verified that this was sufficient to yield
well-converged results.
In most of our runs, we adopted a cosmic ray ionization rate for atomic hydrogen
given by $\zeta_{\rm H} = \zeta_{\rm H, 0} = 10^{-17} \: {\rm s^{-1}}$. However, we 
also carried out a few simulations with larger values of $\zeta_{\rm H}$. Cosmic
ray ionization rates for H$_{2}$, CO etc.\ were computed by scaling the H ionization
rates, as described in \citet{gc12b}.

\begin{table}
\caption{List of simulations \label{tab:sims}}
\begin{tabular}{lcccl}
\hline
ID & ${\rm Z / Z_{\odot}}$ & $G_{0}$ & $\zeta_{\rm H} / \zeta_{\rm H, 0}$ & Notes \\
\hline
Z1-G1 & 1.0 & 1 & 1 & \\
Z1-G10 & 1.0 & 10 & 1 & \\
Z1-G100 & 1.0 & 100 & 1& \\
\hline
Z05-G1 & 0.5 & 1 & 1 & \\
Z05-G10 & 0.5 & 10 & 1 & \\
Z05-G100 & 0.5 & 100 & 1 & \\
\hline
Z03-G1 & 0.3 & 1 & 1 & \\
Z03-G10 & 0.3 & 10 & 1 & \\
Z03-G100 & 0.3 & 100 & 1 & \\
\hline
Z02-G1 & 0.2 & 1 & 1 &\\
Z02-G10 & 0.2 & 10 & 1 & \\
Z02-G100 & 0.2 & 100 & 1 & \\
\hline
Z01-G1 & 0.1 & 1 & 1 & \\
Z01-G10 & 0.1 & 10 & 1 & \\
Z01-G100 & 0.1 & 100 & 1 & \\
\hline
Z003-G1 & 0.03 & 1 & 1 & \\
Z003-G10 & 0.03 & 10 & 1 & \\
Z003-G100 & 0.03 & 100 & 1 & \\
\hline
Z05-G1-D0375 & 0.5 & 1 & 1 & ${\cal D} = 0.375 \, {\cal D_{\odot}}$ \\
Z02-G1-D01 & 0.2 & 1 & 1 &  ${\cal D} = 0.1 \, {\cal D_{\odot}}$ \\
Z01-G1-D001 & 0.1 & 1 & 1 & ${\cal D} = 0.01 \, {\cal D_{\odot}}$ \\
\hline
Z1-G10-CR10 & 1.0 & 10 & 10 & \\
Z1-G100-CR100 & 1.0 & 100 & 100 & \\
Z03-G10-CR10 & 0.3 & 10 & 10 & \\
Z03-G100-CR100 & 0.3 & 100 & 100 & \\
Z01-G10-CR10 & 0.1 & 10 & 10 & \\
Z01-G100-CR100 & 0.1 & 100 & 100 & \\
\hline
\end{tabular}
\\ ${\cal D}$ denotes the dust-to-gas ratio, with ${\cal D_{\odot}}$ being the
value in the local ISM. $G_{0}$ is the strength of the ISRF relative to
\citet{dr78} and \citet{mmp83}.
\end{table}

\section{Results}
\label{result}

\begin{table}
\caption{Onset of star formation in each simulation \label{tab:tsf}}
\begin{tabular}{lc}
\hline
ID & $t_{\rm SF}$~(Myr) \\
\hline
Z1-G1 & 1.98 \\ 
Z1-G10 & 2.17 \\
Z1-G100 & 2.05 \\
\hline
Z05-G1 & 2.43 \\
Z05-G10 & 2.58 \\
Z05-G100 & 2.79 \\
\hline
Z03-G1 & 2.72 \\
Z03-G10 & 2.86 \\
Z03-G100 & 3.24 \\
\hline
Z02-G1 & 2.81 \\
Z02-G10 & 2.97 \\
Z02-G100 & 3.59 \\
\hline
Z01-G1 & 2.95 \\
Z01-G10 & 3.22  \\
Z01-G100 & 4.25 \\
\hline
Z003-G1 & 3.20 \\
Z003-G10 & 3.61 \\
Z003-G100 & 4.81 \\
\hline
Z05-G1-D0375 & 2.54 \\
Z02-G1-D01 & 2.93 \\
Z01-G1-D001 & 2.97 \\
\hline
Z1-G10-CR10 & 2.17 \\   
Z1-G100-CR100 & 2.03 \\ 
Z03-G10-CR10 & 2.88 \\ 
Z03-G100-CR100 & 2.58 \\ 
Z01-G10-CR10 & 3.25 \\ 
Z01-G100-CR100 & --- \\  
\hline
\end{tabular}

\end{table}

\subsection{The \ci-to-H$_{2}$ conversion factor at the onset of star formation}
\begin{figure} 
\includegraphics[width=0.45\textwidth]{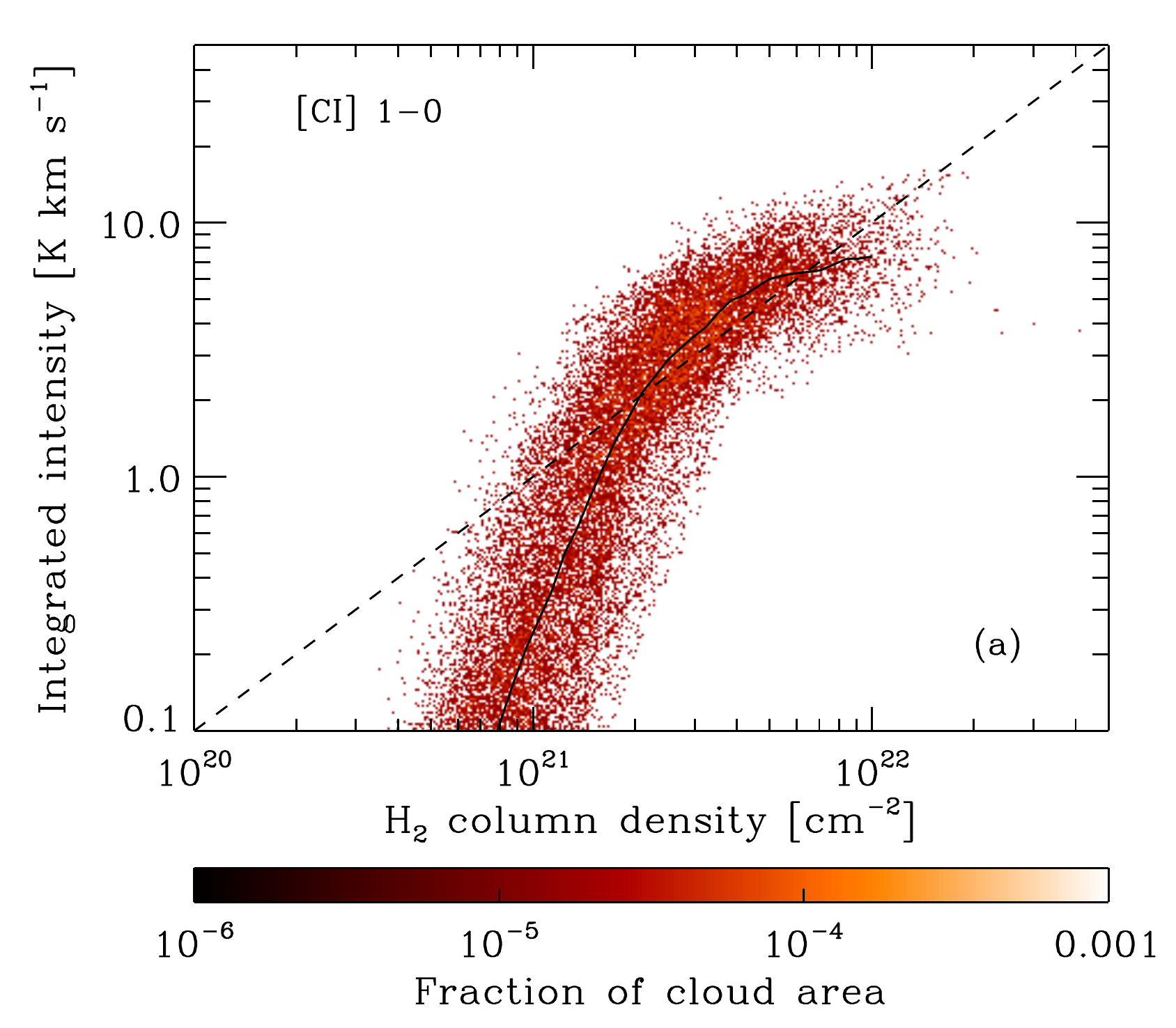}
\includegraphics[width=0.45\textwidth]{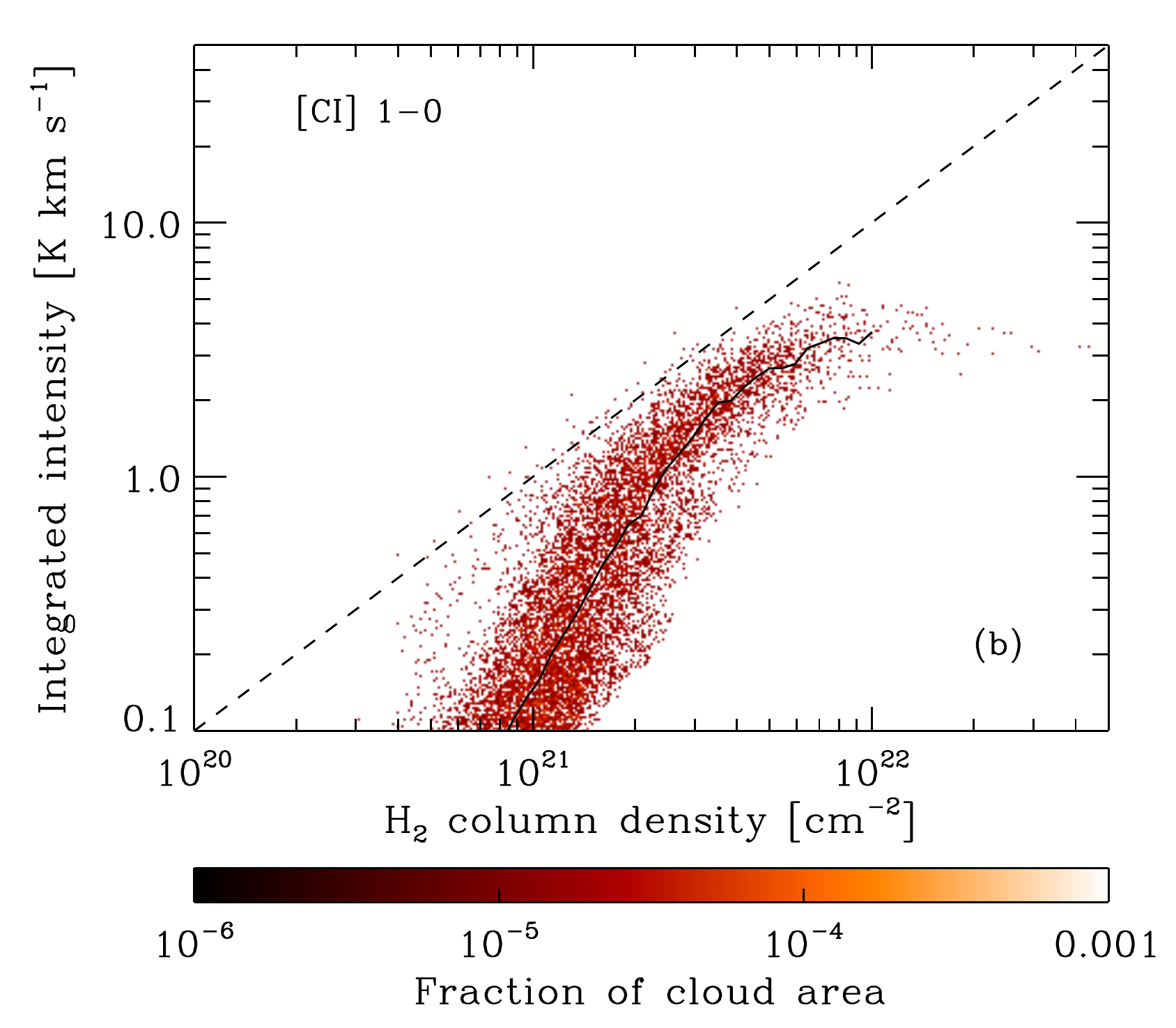}
\caption{(a) Two-dimensional PDF of the integrated intensity of the \ci~$1 \rightarrow 0$ line, plotted
as a function of the H$_{2}$ column density, $N_{\rm H_{2}}$, for run Z1-G10 at $t = t_{\rm SF}$.
The dashed line indicates a linear relationship between $W_{\rm CI, 1-0}$ and $N_{\rm H_{2}}$ and
is included to help guide the eye. The solid line shows the geometric mean of $W_{\rm CI, 1-0}$ as
a function of $N_{\rm H_{2}}$. The points are colour-coded according to the fraction of the total
cloud area that they represent.
(b) As (a), but for run Z02-G1 at $t = t_{\rm SF}$.
\label{fig:WCI_NH2}}
\end{figure}

In Paper I we showed that in a typical turbulent Galactic cloud, the integrated intensity of 
the \ci~$1 \rightarrow 0$ line ($W_{\rm CI, 1-0}$) scales linearly with the H$_{2}$ column density ($N_{\rm H_{2}}$) 
only over a limited range of H$_{2}$ column densities. At high column densities, the correlation
breaks down because the $1 \rightarrow 0$ line becomes optically thick, while at low
column densities, the breakdown reflects the fact that most of the carbon in regions of
low extinction is photoionized to C$^{+}$ by the external radiation field. Observations of
\ci~in Galactic molecular clouds recover similar behaviour \citep[see e.g.][]{beuther14}.
In the present paper, we show that increasing the radiation field strength and/or decreasing
the metallicity does not substantially change this basic conclusion. 
This is illustrated in Figure~\ref{fig:WCI_NH2}, where
we show two examples of the relationship between $W_{\rm CI, 1-0}$ and $N_{\rm H_{2}}$,
taken from runs Z1-G10 and Z02-G1. In both cases, the values shown are for a time immediately
before the onset of star formation, $t = t_{\rm SF}$. The actual physical time that this corresponds to 
for each cloud is listed in Table~\ref{tab:tsf}. We see that in both cases, there is an approximately
linear correlation between $W_{\rm CI, 1-0}$ and $N_{\rm H_{2}}$ only over a relatively narrow 
range in H$_{2}$ column densities. At low column densities, $W_{\rm CI, 1-0}$ falls off rapidly with
decreasing $N_{\rm H_{2}}$, reflecting the fact that most of the carbon is photoionized to C$^{+}$
in regions with low dust extinction. At high column densities, on the other hand, the correlation
between $W_{\rm CI, 1-0}$ and $N_{\rm H_{2}}$ becomes sub-linear because the \ci~$1\rightarrow0$
line becomes optically thick. Moreover, even in the range of H$_{2}$ column densities where the
mean value of  $W_{\rm CI, 1-0}$ does scale approximately linearly with $N_{\rm H_{2}}$, we see
that there is considerable scatter. 

It is clear from this that on small scales, there is no unique
\ci-to-H$_{2}$ conversion factor. Instead, we would expect the conversion factor to vary significantly with 
position within the cloud, similar to the behaviour that we find for $X_{\rm CO}$
\citep[see e.g.][]{pineda08,shetty11a,shetty11b,lee14}.
A more interesting quantity to examine is therefore the mean \ci-to-H$_{2}$ conversion factor for the entire 
cloud. We define this as
 \begin{equation}
 X_{\rm CI} \equiv \frac{\langle N_{\rm H_{2}} \rangle}{\langle W_{\rm CI, 1 \rightarrow 0} \rangle},
 \end{equation}
where the angle brackets denote the fact that we are averaging over the projected spatial extent
of the cloud. We define $X_{\rm CI}$ for the cloud in this way, rather than by computing its value
for each individual pixel and then averaging, for two reasons. First, this better reflects what we
are actually doing when we make observations of unresolved molecular clouds and hence 
measure only the total emission rather than its spatial distribution. Second, this procedure allows
us to deal sensibly with lines-of-sight that have H$_{2}$ without any significant associated  
\ci~emission.

When computing $\langle N_{\rm H_{2}} \rangle$ and $\langle W_{\rm CI, 1 \rightarrow 0} \rangle$,
we include all of the pixels in our projections that have non-zero H$_{2}$ column densities. However,
we note that this is not the only possible way in which to compute `average' H$_{2}$ column densities 
and \ci~integrated intensities for the simulated clouds. A commonly-used alternative method involves
defining some minimum column density or intensity threshold, and then only averaging over the set
of pixels with values above this threshold (see e.g.\ \citealt{cg15}, or Sz\"ucs et~al., in prep.) This
procedure better matches what is done in resolved observations of clouds \citep[e.g.][]{lee14}, while
the procedure that we adopt is more appropriate if the clouds are unresolved.

\begin{figure}
\includegraphics[width=0.45\textwidth]{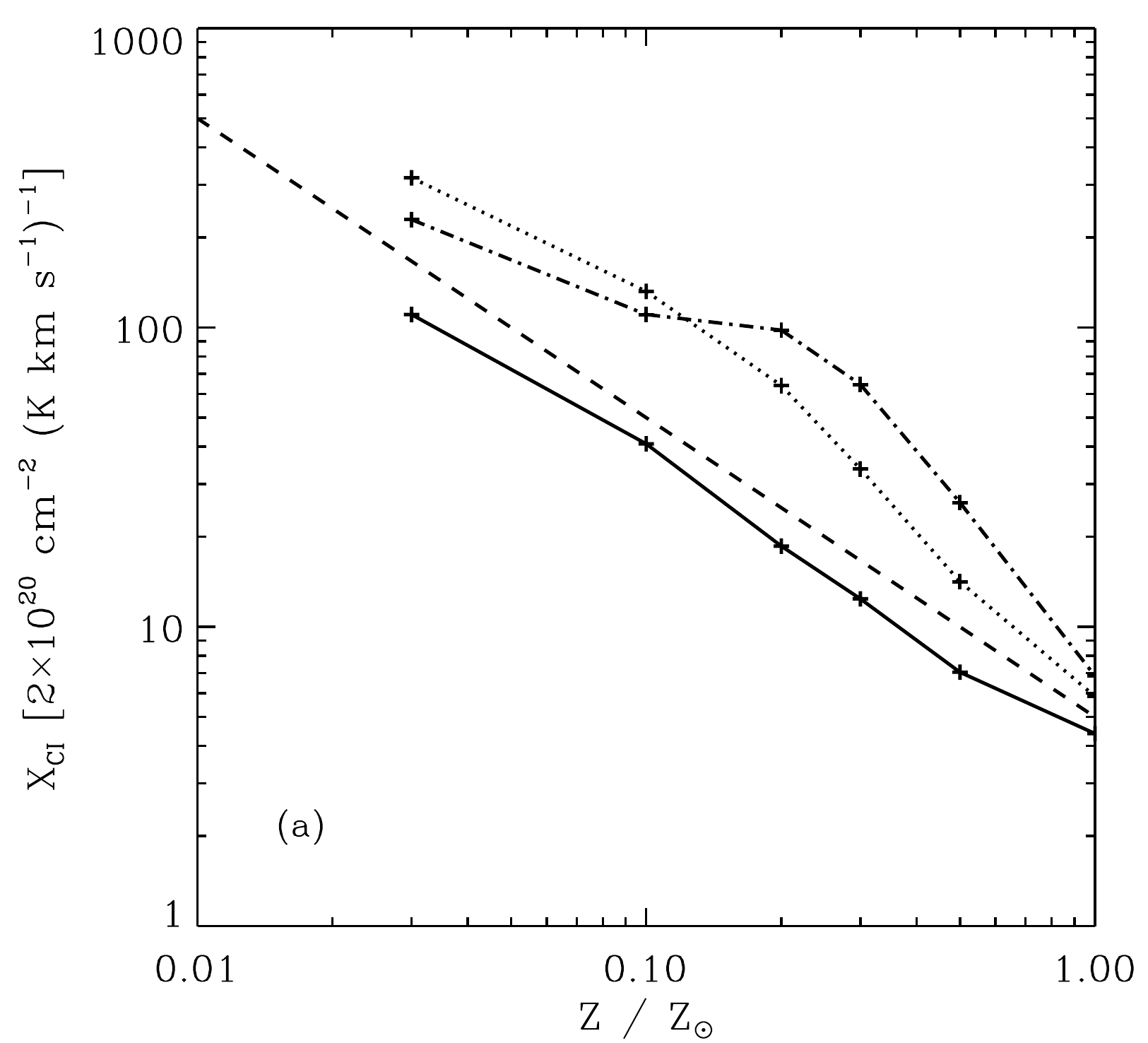}
\includegraphics[width=0.45\textwidth]{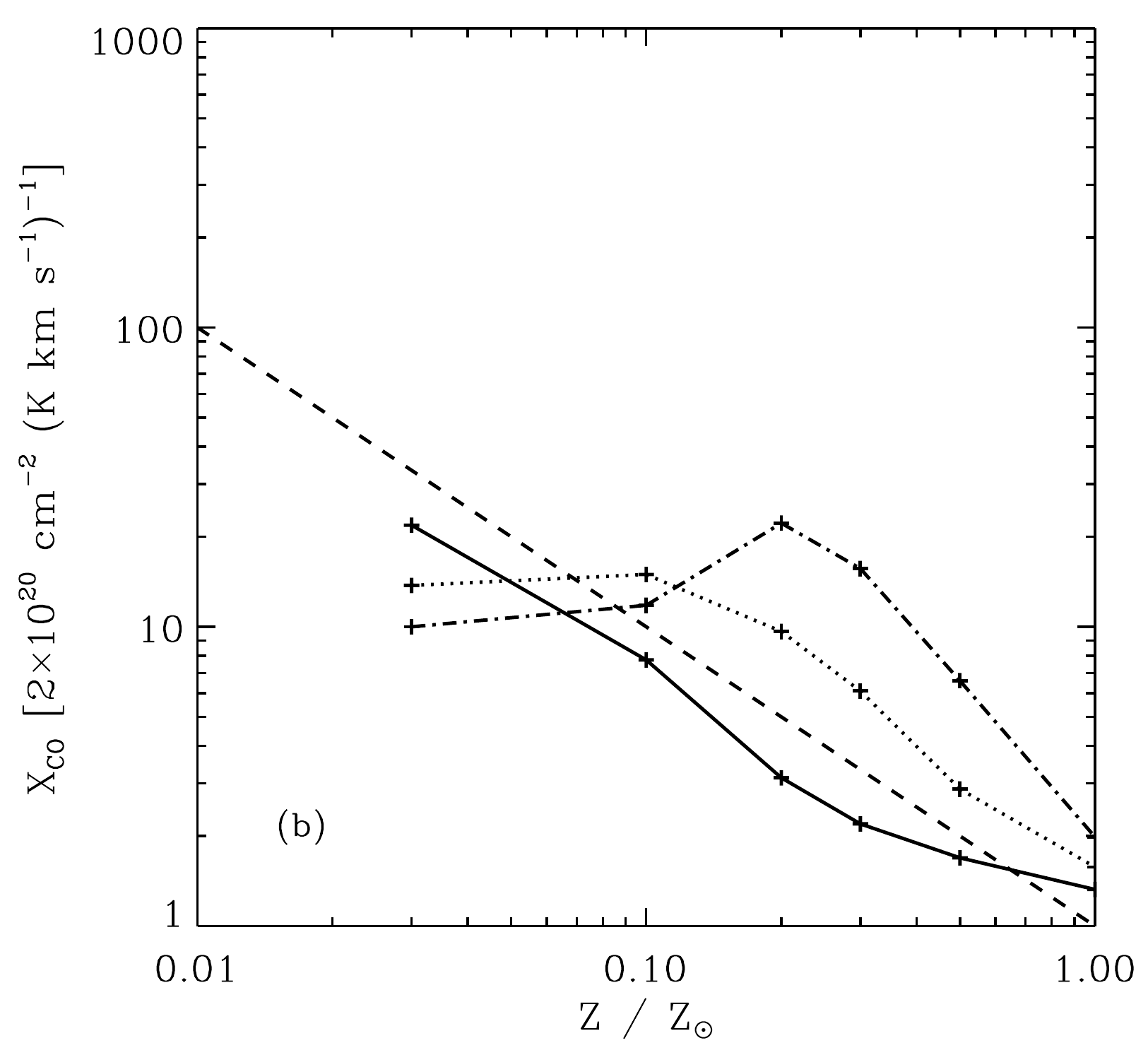}
\caption{(a) \ci-to-H$_{2}$ conversion factor at $t = t_{\rm SF}$, computed for the
\ci~609~$\mu$m line, and plotted in units of the Galactic CO-to-H$_{2}$ conversion factor
($X_{\rm CO, MW} = 2 \times 10^{20} \: {\rm cm^{-2} (K \, km/s)^{-1}}$) for a range of 
different metallicities and radiation field strengths. 
We show results for radiation field strengths $G_{0} = 1$ (solid line), $G_{0} = 10$ (dotted line)
and $G_{0} = 100$ (dash-dotted line). The dashed line indicates a slope of ${\rm Z}^{-1}$ and
is included merely to guide the eye.
(b) As (a), but for the CO-to-H$_{2}$ conversion factor,
$X_{\rm CO}$. \label{fig:XCI_Z}}
\end{figure}

In Figure~\ref{fig:XCI_Z}a, we show how $X_{\rm CI}$ varies as a function of metallicity 
and radiation field strength for our simulated clouds. Once again, we show the results for
$t = t_{\rm SF}$. The values of $X_{\rm CI}$ shown in this figure are given in terms of the
canonical Galactic CO-to-H$_{2}$ conversion factor, $X_{\rm CO, MW}$.

Two points are immediately obvious. First, at solar metallicity, $X_{\rm CI}$ is significantly
larger than $X_{\rm CO, MW}$. In run Z1-G1, we recover a value of 
$X_{\rm CI} = 8.8 \times 10^{20} \: {\rm cm^{-2} \: K^{-1} \: km^{-1} \: s}$, 
within 25\% of the value computed by \citet{offner14} for a somewhat denser cloud. This is 
more than a factor of 4 larger than $X_{\rm CO, MW}$, and roughly three times larger than the 
value of $X_{\rm CO}$ that we recover for the same simulated cloud (see Figure~\ref{fig:XCI_Z}b). 
Increasing the strength of the ISRF by a factor of 10--100 increases $X_{\rm CI}$ by around 
30--50\%, but also increases $X_{\rm CO}$ by roughly the same amount. 
Therefore, at solar metallicity and at this point in the clouds' evolution, there does not appear
to be a compelling reason to prefer \ci~over CO as a tracer of the total cloud mass.

Second, Figure~\ref{fig:XCI_Z}a demonstrates that $X_{\rm CI}$ systematically increases 
as we reduce the metallicity of the gas, with a scaling that is close to $X_{\rm CI} \propto
{\rm Z}^{-1}$ in the $G_{0} = 1$ runs. In the runs with stronger radiation fields, we see a
slightly steeper scaling for metallicities in the range $0.2 < {\rm Z / Z_{\odot}} < 1.0$, 
which then flattens as we move to even lower metallicities.  

\begin{figure}
\includegraphics[width=0.5\textwidth]{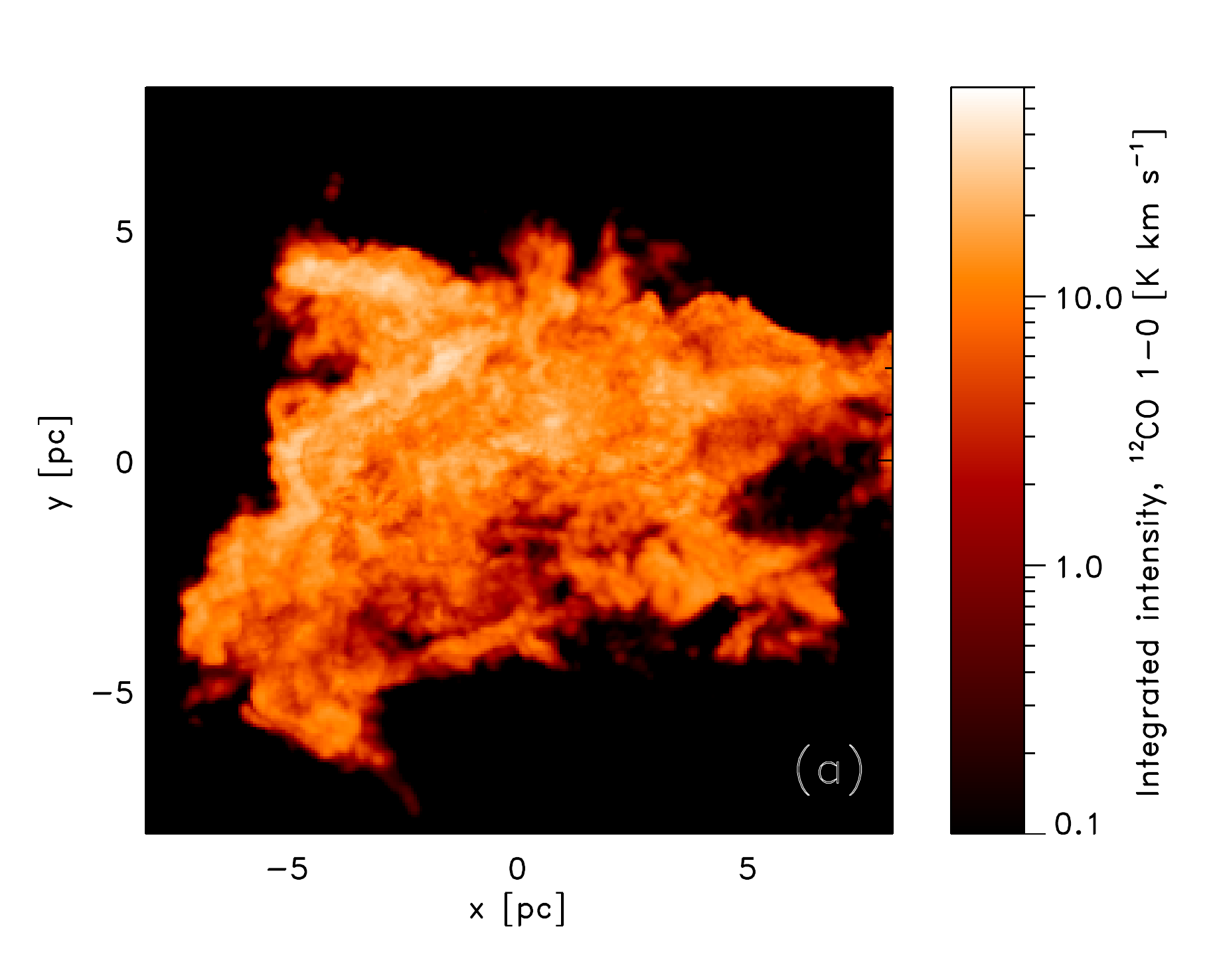}
\includegraphics[width=0.5\textwidth]{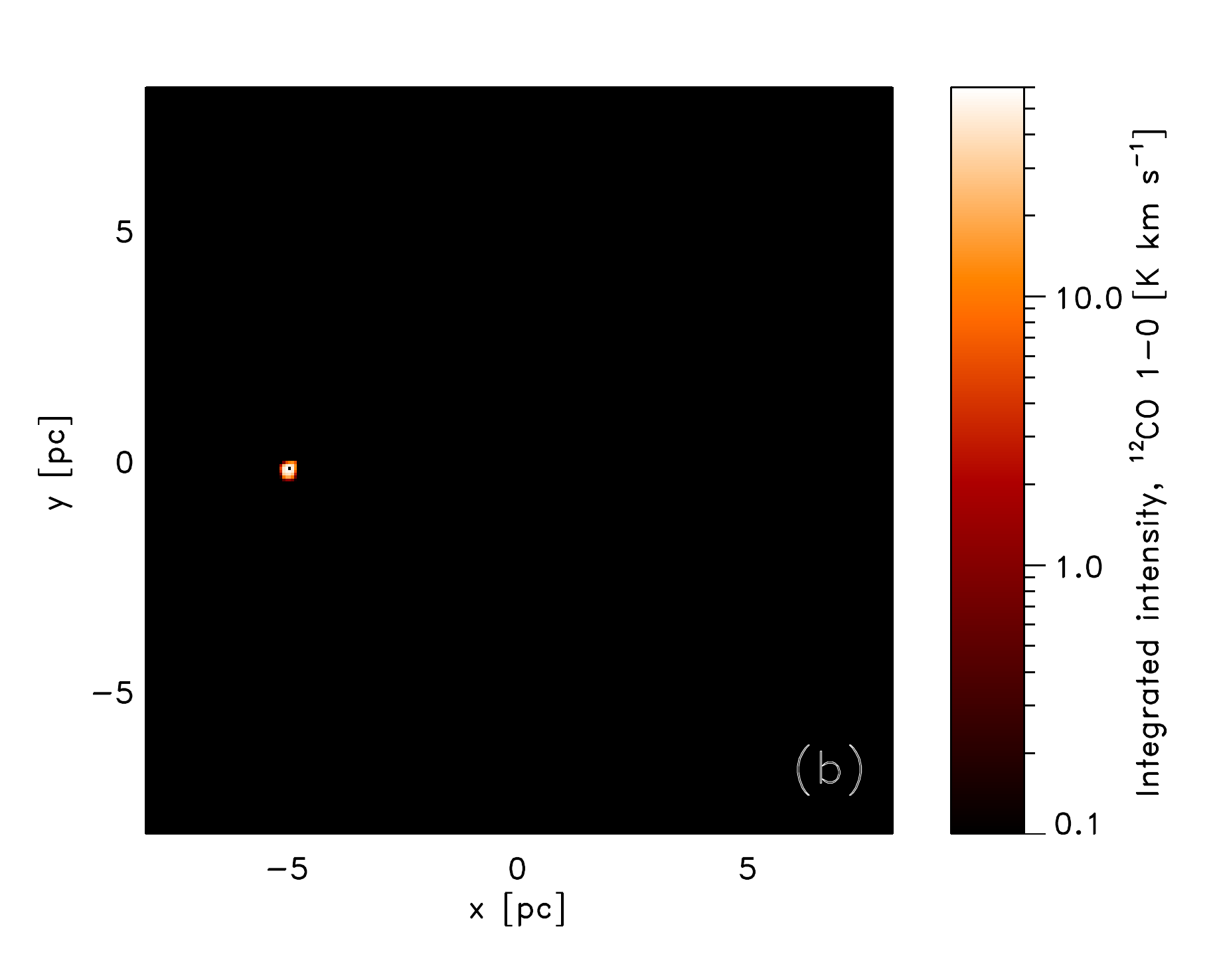}
\caption{(a) Map of the integrated intensity of the $J = 1 \rightarrow 0$ 
transition of $^{12}$CO in run Z1-G1 at time $t = t_{\rm SF}$.
(b) As (a), but for run Z01-G100.
\label{fig:map}} 
\end{figure}

For comparison, we illustrate in Figure~\ref{fig:XCI_Z}b how the value of $X_{\rm CO}$
that we measure at the onset of star formation varies with metallicity. As in the case of
$X_{\rm CI}$, we see that at ${\rm Z} \geq 0.2 \: {\rm Z_{\odot}}$, $X_{\rm CO}$ increases
as the metallicity decreases, with a dependence that again scales roughly as  Z$^{-1}$.
At lower metallicities, the behaviour becomes more complicated. When $G_{0} = 1$,
$X_{\rm CO}$ continues to increase with decreasing metallicity, but in the $G_{0} = 10$
and $G_{0} = 100$ runs we find instead that $X_{\rm CO}$ remains constant or even
decreases with decreasing Z. This occurs because in these conditions, the total CO
emission of the cloud becomes dominated by the contribution from a single dense,
compact, gravitationally-collapsing clump (see Figure~\ref{fig:map}). The high column density of the 
gas in this clump provides effective shielding for the CO even when the metallicity is very low, and 
allows  the gas to build up a large CO column density within the clump. Consequently, the CO
emission from the clump becomes optically thick, at which point it becomes only
weakly sensitive to the fractional abundance of CO within the dense gas. 
As a result, the mean value of the $^{12}$CO integrated intensity, $\langle W_{\rm CO} \rangle$, 
which varies strongly with metallicity at
high ${\rm Z}$, starts to vary only weakly with metallicity once we reach this clump-dominated
regime. At the same time, however, the H$_{2}$ content of the cloud still decreases 
substantially with decreasing metallicity even once we reach this clump-dominated regime,
since even at the lowest metallicities, much of the H$_{2}$ is still located outside of the
dense clump. 
  
\begin{figure}
\includegraphics[width=0.45\textwidth]{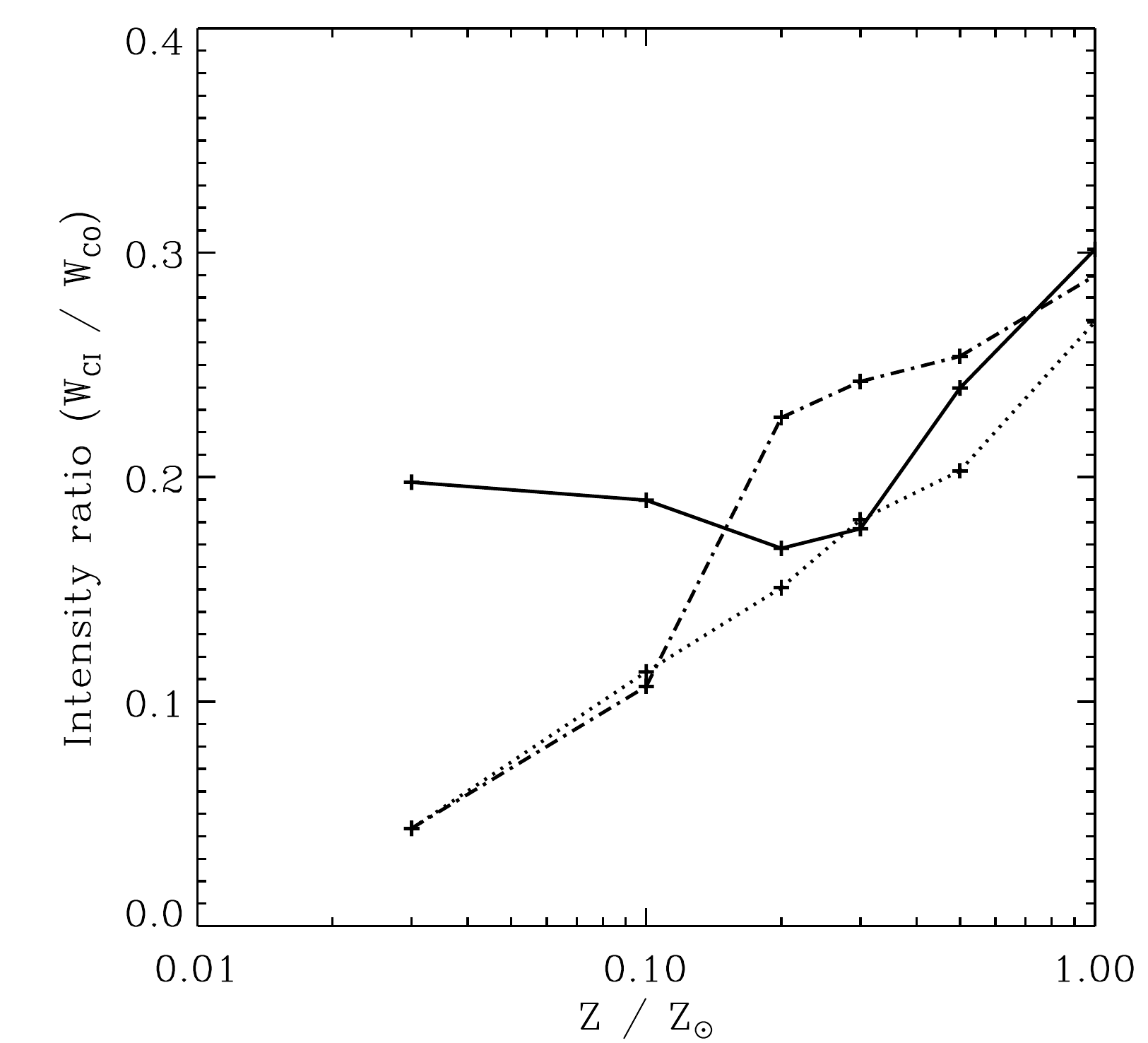}
\caption{Ratio of the mean integrated intensities (in K~km~s$^{-1}$)
of the \ci~609~$\mu$m line 
($\langle W_{\rm CI} \rangle$) and the $^{12}$CO $J = 1 \rightarrow 0$ line 
($\langle W_{\rm CO} \rangle$), computed at $t = t_{\rm SF}$ and plotted as 
a function of metallicity.  We show results 
for radiation field strengths $G_{0} = 1$ (solid line), $G_{0} = 10$ (dotted line)
and $G_{0} = 100$ (dash-dotted line). \label{fig:WCI_WCO}}
\end{figure}

The results of our comparison show that at this point in the evolution of the cloud, 
the cloud-averaged CO emission remains a better tracer of the molecular mass 
than the cloud-averaged \ci~emission, even at metallicities as low as 
${\rm Z} = 0.03 \: {\rm Z_{\odot}}$. This can be seen more easily if we plot the
ratio of $X_{\rm CO}$ to $X_{\rm CI}$, which is the same as the ratio of the
cloud-averaged \ci~and  CO integrated intensities. This is shown in Figure~\ref{fig:WCI_WCO}.
We see that in clouds at the onset of star formation, $\langle W_{\rm CI} \rangle / \langle W_{\rm CO} \rangle$
decreases as we decrease the metallicity, although the dependence is fairly
weak: an order of magnitude drop in Z leads to a decrease in the intensity ratio
of no more than a factor of two. We also see that except at the very lowest 
metallicity that we study,  $\langle W_{\rm CI} \rangle / \langle W_{\rm CO} \rangle$ 
is highly insensitive to the value of $G_{0}$.

\subsection{Time evolution of $X_{\rm CI}$ and $X_{\rm CO}$}
\label{sec:time_evol}
In the previous section, we examined the behaviour of $X_{\rm CI}$ and
$X_{\rm CO}$ in the case where the gas has already collapsed and started 
forming stars, and showed that in this case, CO is a superior tracer of the
molecular mass, at least for the range of metallicities and UV field strengths
considered here. However, it does not immediately follow that CO will be
a superior tracer of H$_{2}$ at all times during the evolution of the cloud.
As we have already seen, at low metallicity the bulk of the CO emission
produced by the cloud comes from between one and a few dense, compact,
collapsing clumps (see also \citealt{gc12b}, who report similar results).
These clumps have higher volume
densities and higher extinctions than the bulk of the cloud, and hence are
the regions in which CO is best able to resist photodissociation. However,
the dynamical timescale on which these clumps evolve is relatively short:
they will either collapse on a free-fall timescale ($t_{\rm ff} \simeq 0.5$~Myr
for clumps with a mean density $n \sim 10^{4} \: {\rm cm^{-3}}$, or even 
shorter at higher densities), or will be disrupted by turbulence on a similar
timescale. This prompts us to ask how $X_{\rm CI}$
and $X_{\rm CO}$ evolve as a function of time: do we find larger values at 
earlier times, when these dense clumps have not yet formed?

\begin{figure} 
\includegraphics[width=0.45\textwidth]{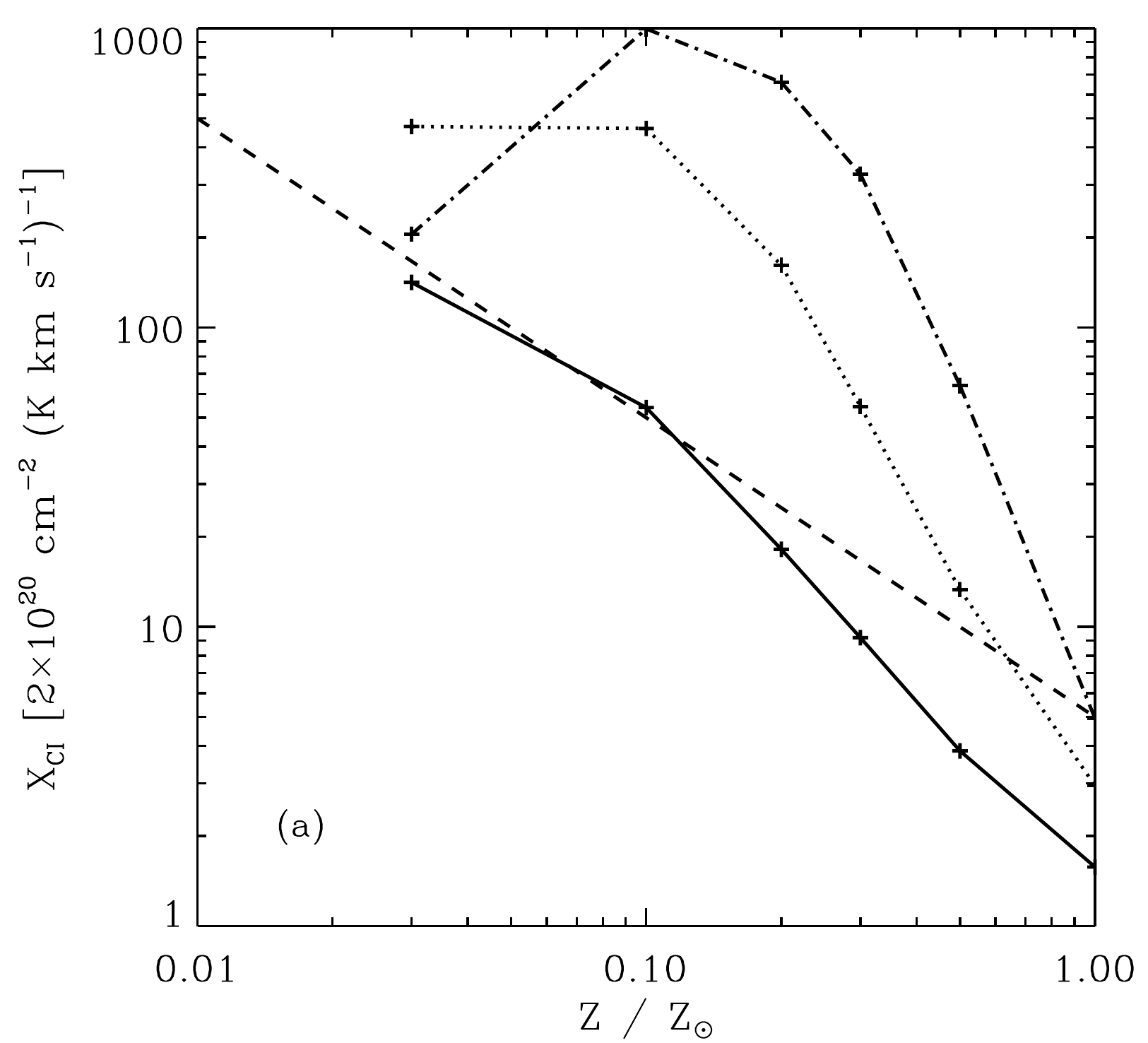}
\includegraphics[width=0.45\textwidth]{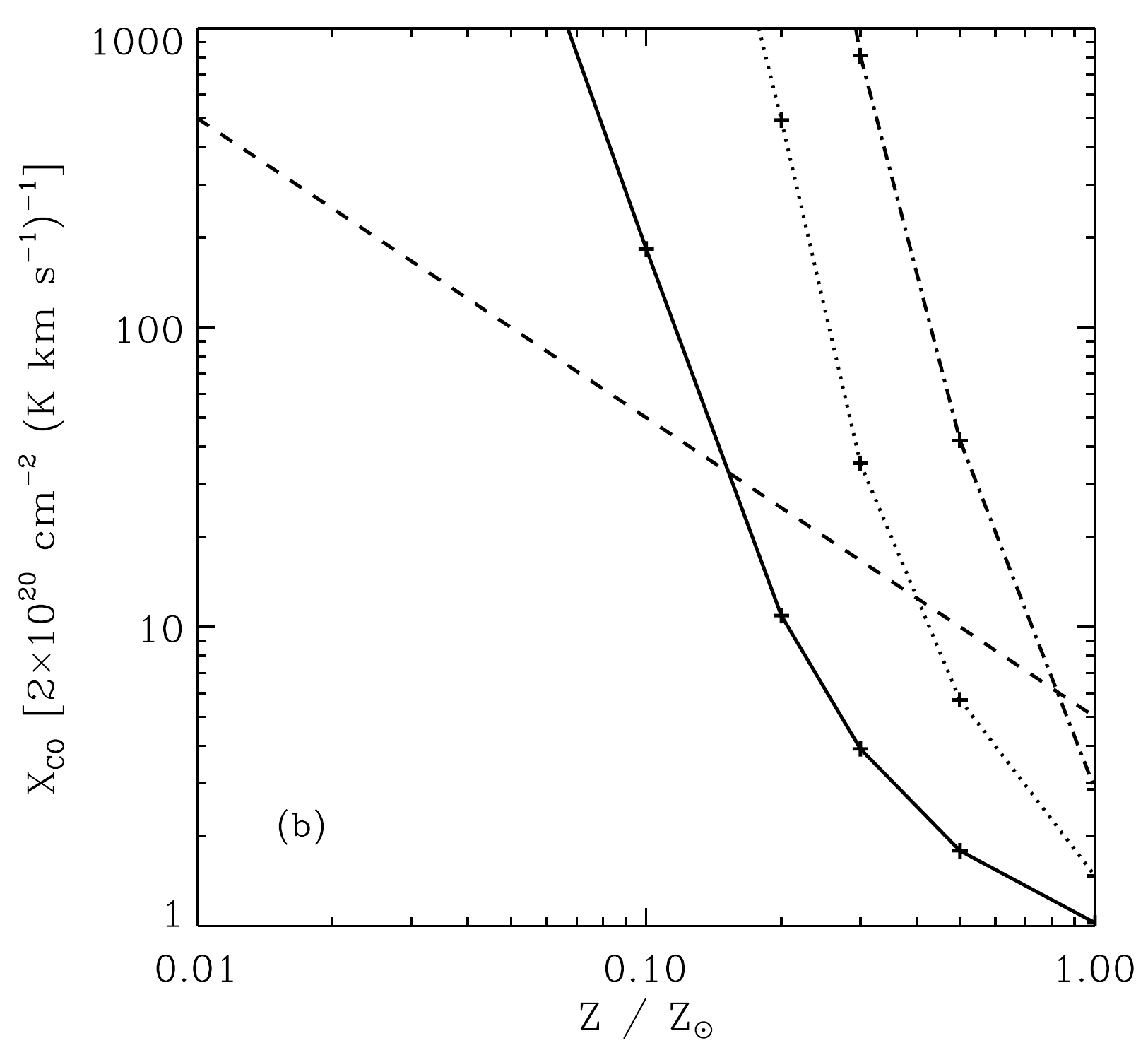}
\caption{(a) As Figure~\ref{fig:XCI_Z}a, but showing values of
$X_{\rm CI}$  computed at $t = t_{\rm SF} - 1 \: {\rm Myr}$.
(b) As (a), but for $X_{\rm CO}$.
\label{fig:XCI_Z_t1}}
\end{figure}

To investigate this, we have calculated $X_{\rm CI}$ and $X_{\rm CO}$
for each of our clouds at a time $t = t_{\rm SF} - 1$~Myr. The results are
plotted in Figure~\ref{fig:XCI_Z_t1}. 
We see that when $G_{0} = 1$, the values of $X_{\rm CI}$ that we measure at
$t = t_{\rm SF} - 1$~Myr are similar to or smaller than those that we measure at 
$t = t_{\rm SF}$. In other words, in these conditions, \ci~emission remains a 
good tracer of the molecular mass of the cloud even prior to the onset of 
gravitational collapse. The same is true in the runs with $G_{0} = 10$ and 100
at solar metallicity, but not at lower metallicities, where we find a much steeper
increase in $X_{\rm CI}$ with decreasing Z than we did in the previous section.
That said, the actual increase in the values of $X_{\rm CI}$ relative to those
in the $t = t_{\rm SF}$ case remains relatively modest: at most, we see increases
of around a factor of a few.

The story is very different in the case of CO. At solar metallicity, much of
the CO emission comes from gas with a number density of a few
thousand cm$^{-3}$. Regions with this density are found throughout the
cloud \citep[see e.g.\ Figure~5 in][]{gc12b} and most of this gas is not
self-gravitating. Individual CO-bright regions are formed and destroyed
by the action of the turbulence, but the total amount of gas in these regions
does not vary much. Consequently, $X_{\rm CO}$ does not vary strongly with time. 
At much lower metallicities, however, dense, self-gravitating clumps dominate the total 
emission, and the value of $X_{\rm CO}$ becomes highly sensitive to the evolutionary
state of these clumps. In this case, the values that we measure for $X_{\rm CO}$ at 
$t = t_{\rm SF} - 1$~Myr are much, much larger than those at $t = t_{\rm SF}$. 

\begin{figure}
\includegraphics[width=0.45\textwidth]{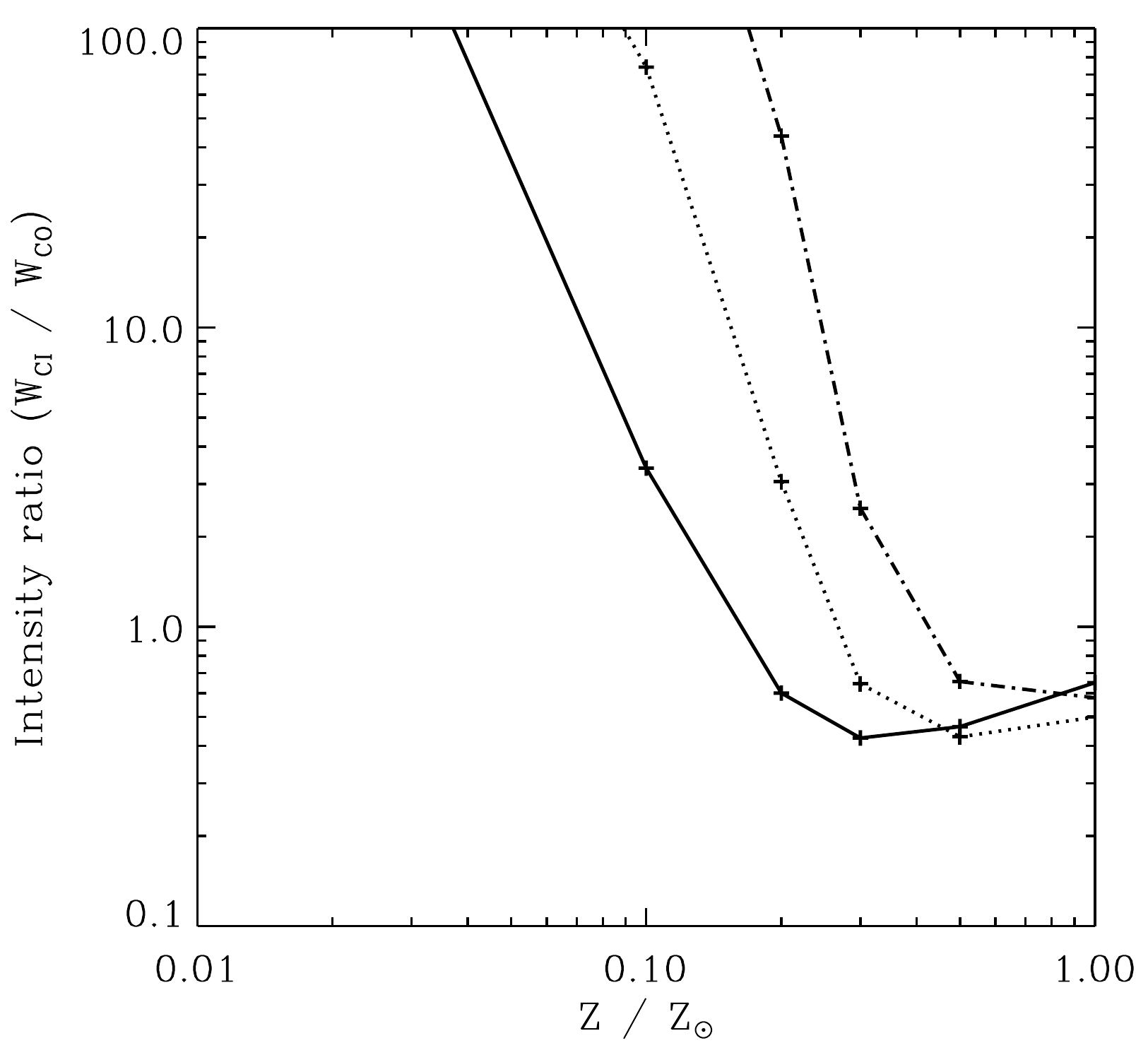}
\caption{As Figure~\ref{fig:WCI_WCO}, but for $t = t_{\rm SF} - 1$~Myr. 
We show results for radiation field strengths $G_{0} = 1$ (solid line), $G_{0} = 10$ (dotted line)
and $G_{0} = 100$ (dash-dotted line). Note the difference in vertical scale
between this figure and Figure~\ref{fig:WCI_WCO}.
\label{fig:WCI_WCO_t1}}
\end{figure}   

This point is further emphasized if we look at the \ci/CO integrated intensity
ratio at the same time (Figure~\ref{fig:WCI_WCO_t1}). At metallicities close
to solar, the ratio of the \ci and CO integrated intensities at this point in the cloud's 
evolution is around 0.5 to 0.6 (for values expressed in terms of K~km~s$^{-1}$), 
somewhat larger than at $t = t_{\rm SF}$, but of the same
order of magnitude. At lower metallicities, however, the ratio increases
dramatically as the diffuse CO emission vanishes. When $G_{0} = 1$, this
sharp increase in the \ci/CO intensity ratio occurs at a metallicity of around
$0.1 \: {\rm Z_{\odot}}$, but increasing $G_{0}$ leads to this change of
behaviour happening at higher metallicity, owing to the more effective
destruction of the diffuse CO in runs with a higher radiation field strength.
     
\begin{figure}
\includegraphics[width=0.45\textwidth]{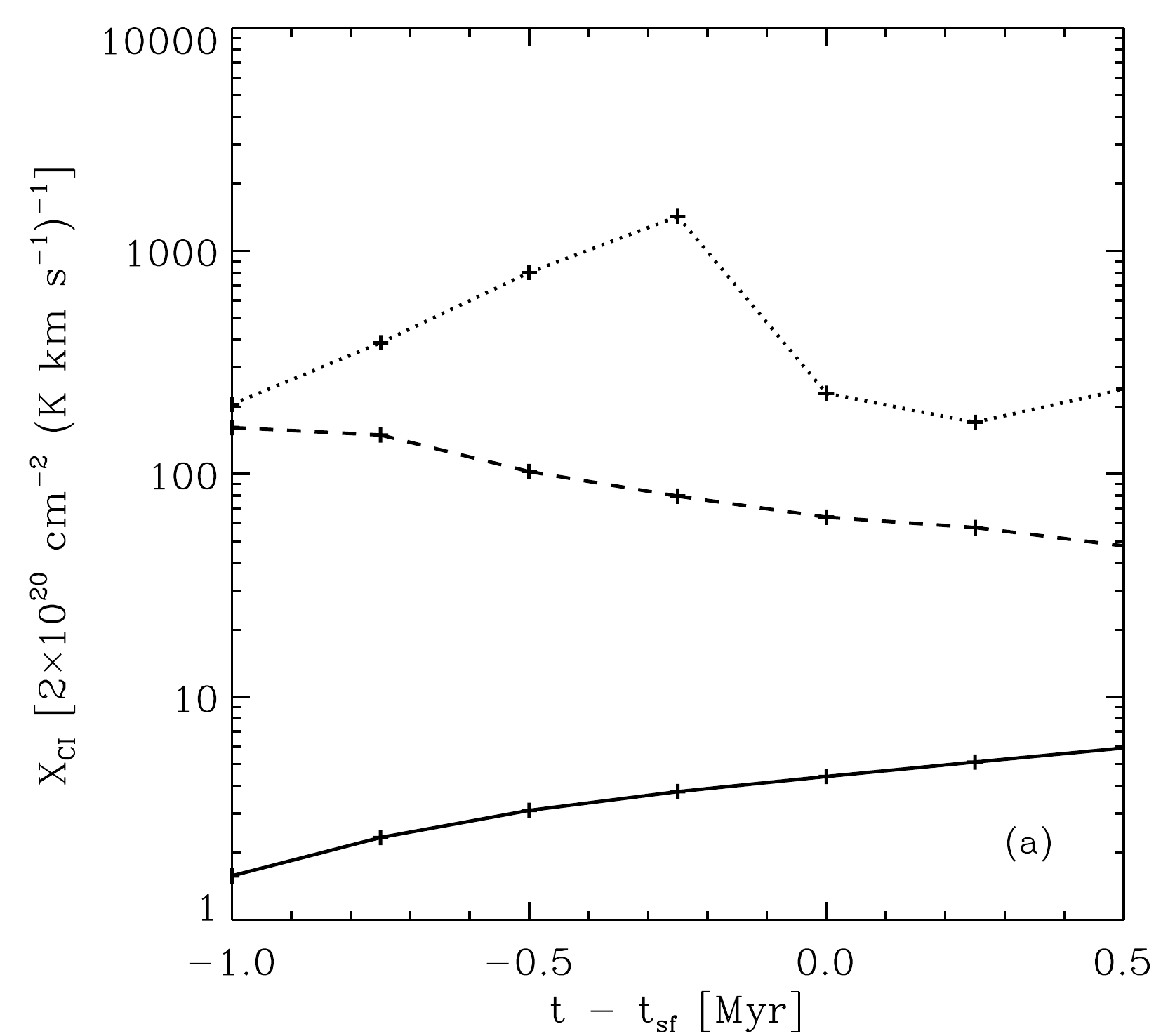}
\includegraphics[width=0.45\textwidth]{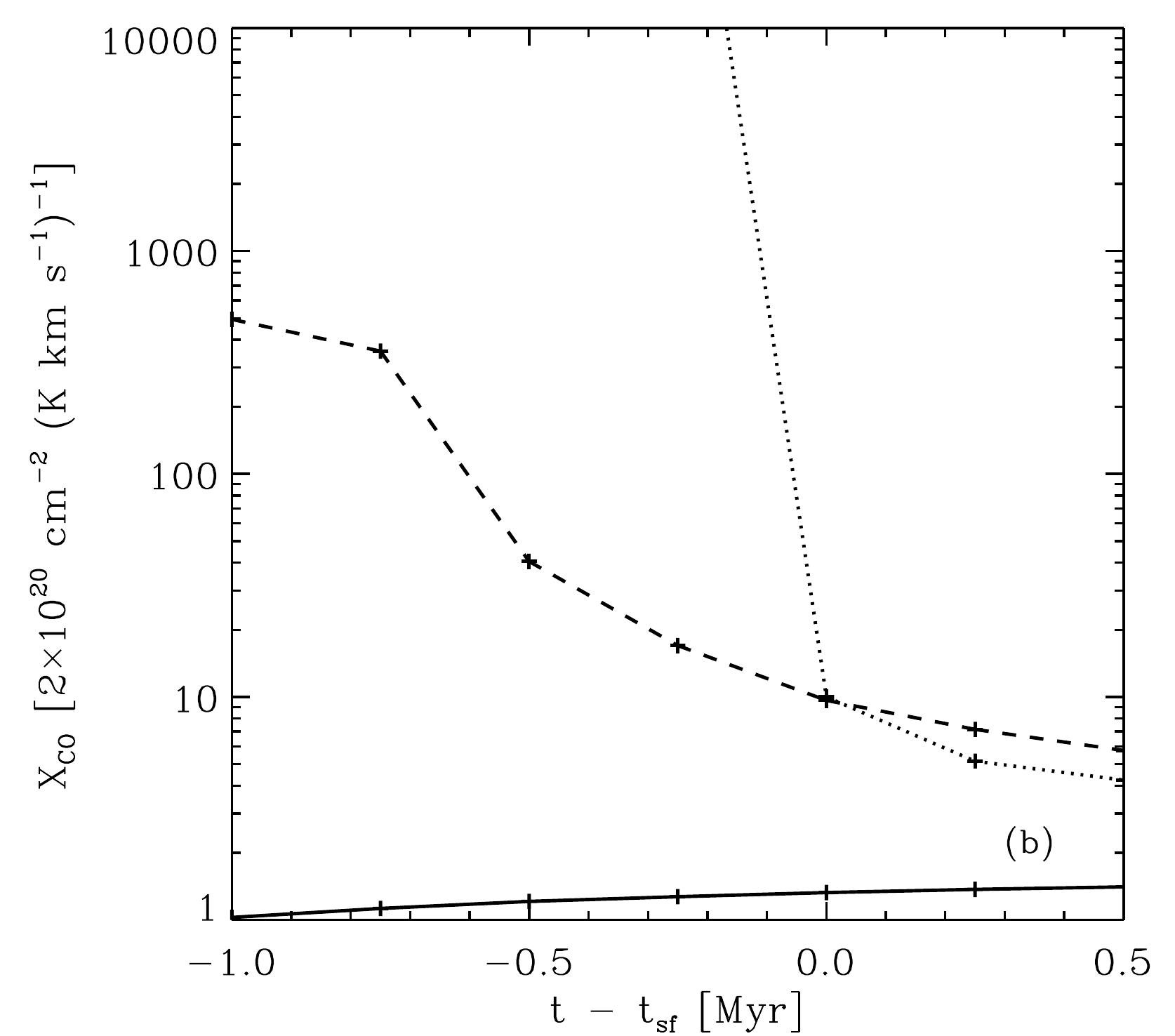}
\caption{(a) Evolution with time of $X_{\rm CI}$ in runs Z1-G1 (solid line),
Z02-G10 (dashed line) and Z003-G100 (dotted line). 
(b) As (a), but showing the evolution with time of $X_{\rm CO}$.
\label{fig:XC_time}}
\end{figure}   

In Figure~\ref{fig:XC_time}a, we explore in more detail how $X_{\rm CI}$ and $X_{\rm CO}$
evolve with time in several representative runs.  In run Z1-G1, $X_{\rm CI}$ rises gradually
over time as the cloud collapses and an increasing fraction of the available carbon is
converted into CO. In run Z02-G10, on the other hand, the collapse of the cloud leads to
a decrease in $X_{\rm CI}$, as in this case, the increased conversion of C$^{+}$ to C is
more important than the conversion of C to CO. Finally, in run Z003-G100, there is no
clear trend in the evolution of $X_{\rm CI}$ with time. Importantly, we see that in no case
does $X_{\rm CI}$ vary by more than a factor of a few over the time period examined,
which corresponds to roughly half a free-fall time. 

If we now look at the evolution of $X_{\rm CO}$ in the same three runs over the same
period of time (Figure~\ref{fig:XC_time}b), we see that the behaviour is very different.
In run Z1-G1, the dense clumps and filaments created by the turbulence provide enough
shielding to allow significant quantities of CO to form even prior to the onset of
run-away gravitational collapse. Consequently, the CO content of the cloud does not
change substantially over time and $X_{\rm CO}$ remains approximately constant.

In run Z02-G10, on the other hand, the combination of the lower dust extinction and
higher radiation field strength mean that the over-dense structures created by the 
turbulence alone contain little CO. In this case, it is the gravitational collapse of
some of these structures that triggers a substantial increase in the CO luminosity
of the cloud. The result is that in this run, $X_{\rm CO}$ evolves strongly over time,
decreasing by almost a factor of 100 over the 1.5~Myr period shown here. 

Finally, we see that in run Z003-G100, the time-dependence of $X_{\rm CO}$ becomes
even more extreme. In this run, CO forms only at very high densities, shortly before
the onset of star formation, and so $X_{\rm CO}$ rapidly changes from being orders of 
magnitude larger than the Galactic value at $t \ll t_{\rm SF}$ to only 5--10 times larger
than the Galactic value at $t \sim t_{\rm SF}$ and later times.
     
\subsection{Dependence of $W_{\rm CI}$ and $W_{\rm CO}$ on extinction}
\begin{figure}
\includegraphics[width=0.45\textwidth]{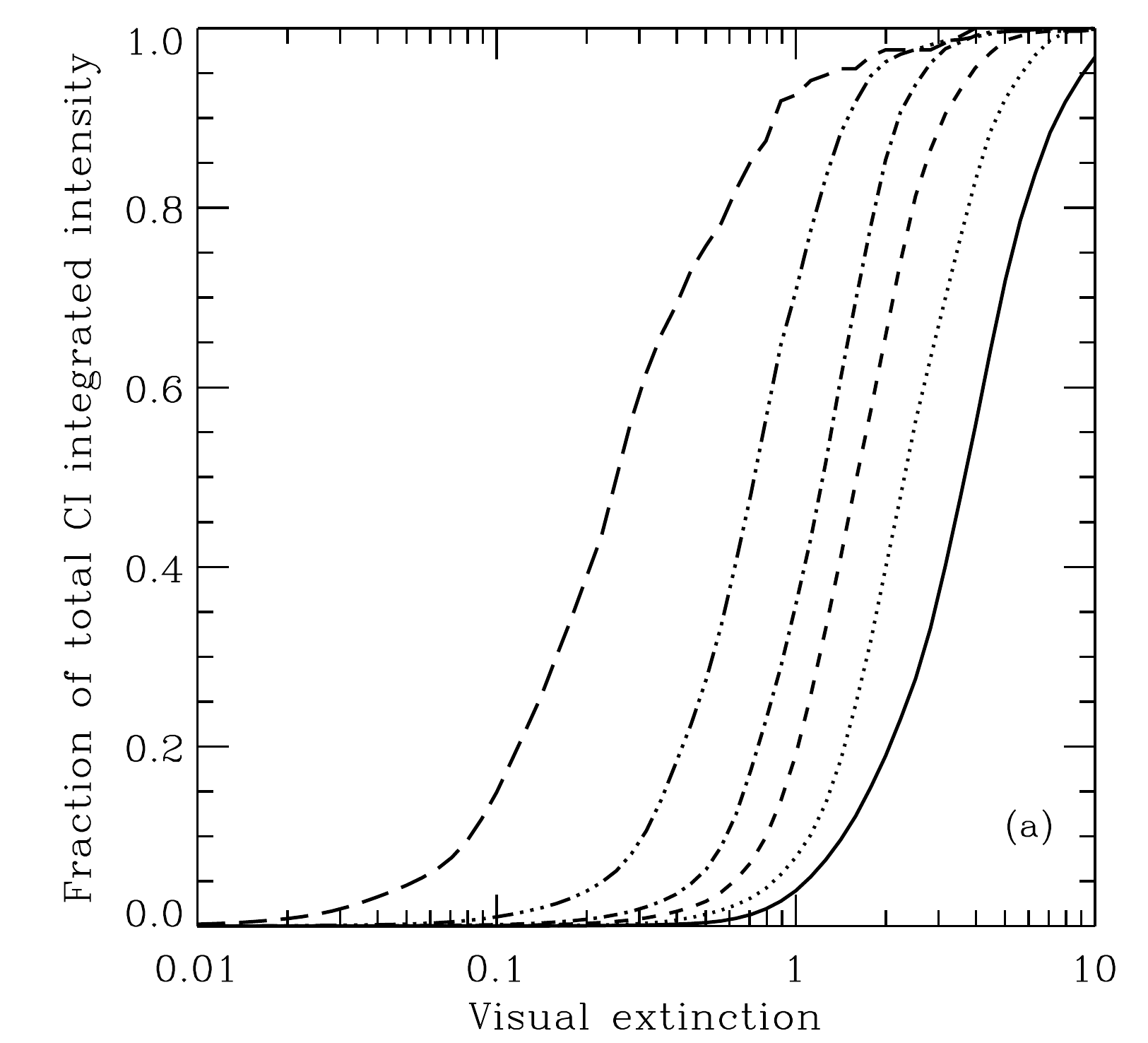}
\includegraphics[width=0.45\textwidth]{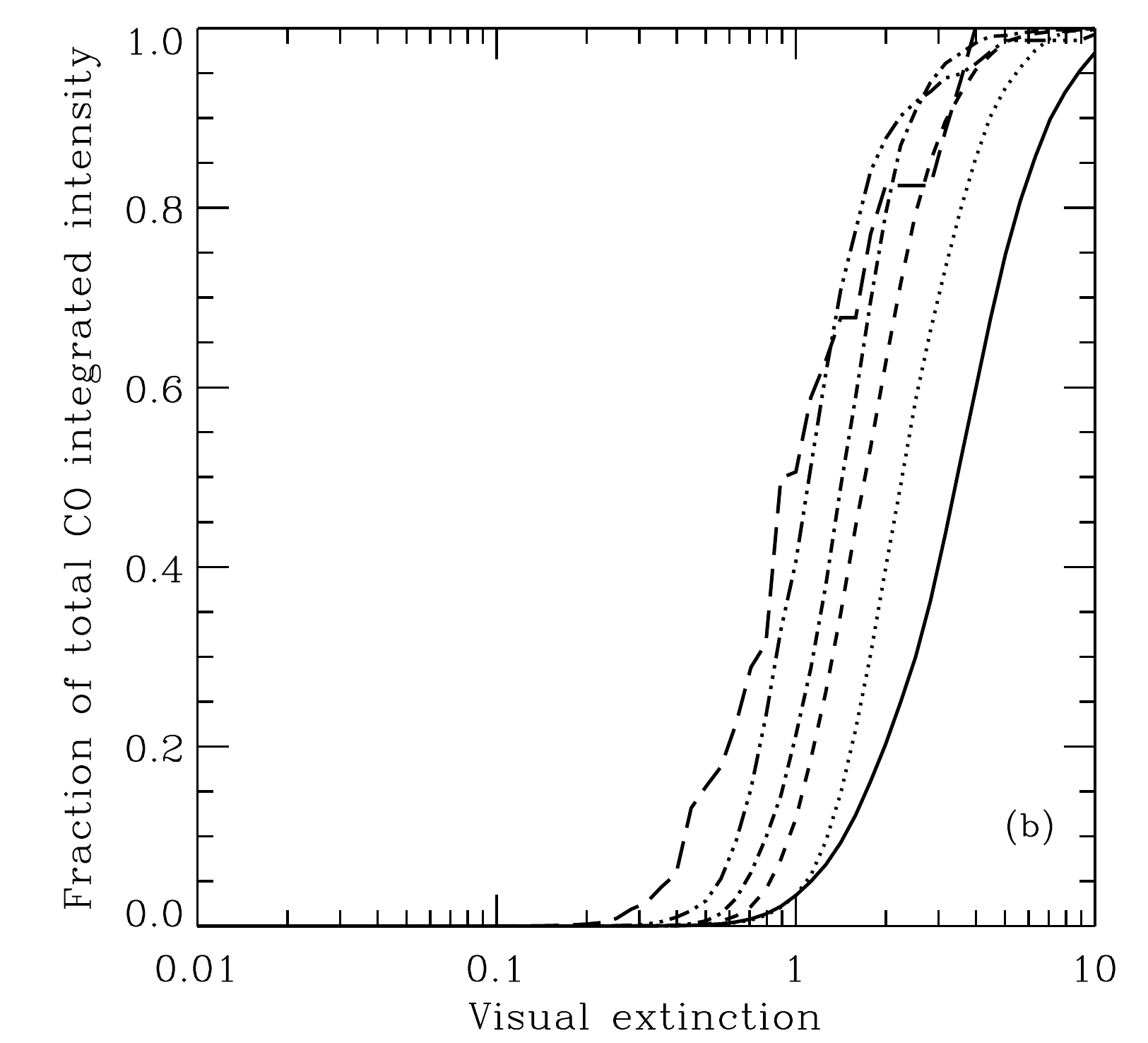}
\caption{(a) Fraction of the total \ci~emission from the simulated clouds coming from
lines of sight with values of $A_{\rm V}$ at or below the specified value, measured
at $t = t_{\rm SF}$. Results are
plotted for runs Z1-G1 (solid), Z05-G1 (dotted), Z03-G1 (dashed), Z02-G1 (dot-dashed),
Z01-G1 (dot-dot-dot-dashed) and Z003-G1 (long-dashed). 
(b) As (a), but for the total $^{12}$CO $J = 1 \rightarrow 0$ emission.
\label{fig:WCI_AV}}
\end{figure}

To help us understand why $X_{\rm CI}$ is much less time dependent than $X_{\rm CO}$
in low metallicity clouds, it is useful to look at how the integrated intensities of the
\ci~609$\,\mu$m line and the CO $J = 1 \rightarrow 0$ line depend on the line-of-sight
visual extinction of the gas and how this dependence changes
as we vary the metallicity. In Figure~\ref{fig:WCI_AV}a, we show a cumulative
plot of the fraction of the total \ci~emission coming from lines-of-sight at or below the
specified visual extinction for a number of different simulations that cover the full range
of metallicities that we examine in this paper. In each case, the curves
correspond to the state of the cloud at $t = t_{\rm SF}$. In Figure~\ref{fig:WCI_AV}b,
we show a similar plot for the CO $J = 1 \rightarrow 0$ emission in the same set of
simulations.

We see in the solar metallicity run, almost all of the \ci~emission is confined to lines-of-sight
with $1 < A_{\rm V} < 10$, with roughly 50\% coming from the narrower range 
$2.5 < A_{\rm V} < 5$. However, as we decrease the metallicity, the range of extinctions
occupied by the lines-of-sight producing most of the \ci~emission shifts to systematically
lower values, so that by the time we reach the $0.03 \: {\rm Z_{\odot}}$ run, almost all of
the emission is coming from gas with $A_{\rm V} < 1$.  The amount by which the curve
shifts from run to run roughly corresponds to the change in the dust-to-gas ratio, which
is consistent with the same range of gas column densities dominating the emission in
each case, regardless of the dust-to-gas ratio. This range of column densities extends
from a few times $10^{21} \: {\rm cm^{-2}}$ to $\sim 10^{22} \: {\rm cm^{-2}}$, and it is
important to note that the mean H nuclei column density of the cloud at $t = 0$, which is approximately
$\langle N_{\rm H} \rangle_{\rm init} \simeq 7 \times 10^{21} \: {\rm cm^{-2}}$, lies in the middle of
this range. It is therefore not surprising that the clouds are bright in \ci~right from the start
of the simulation, and hence that $X_{\rm CI}$ does not vary strongly with time. 

In the case of CO, however, we see quite different behaviour, as Figure~\ref{fig:WCI_AV}b
demonstrates. In run Z1-G1, most of the CO emission is produced by lines of sight with
$2 < A_{\rm V} < 5$, similar to the behaviour of the \ci~emission. As we decrease the
metallicity, we again find that the range of extinctions responsible for producing most of
the emission shifts to lower values. However, in this case, the lines shift by a much smaller
amount: a factor of 30 change in metallicity and dust-to-gas ratio shifts the curves by no
more than around a factor of three. This means that the range of gas column densities
that dominate the total CO emission shifts to higher values as we decrease the metallicity.
For example, in run Z01-G1, around 50\% of the total CO $J = 1 \rightarrow 0$ emission
is produced by lines-of-sight with $A_{\rm V} > 1$, corresponding to $N_{\rm H} \simeq
2 \times 10^{22} \: {\rm cm^{-2}}$. 

This behaviour provides a simple explanation for the strong time dependence that we
see in $X_{\rm CO}$ at low metallicities. In our low metallicity clouds, the column density
required to provide sufficient dust shielding to allow CO to form in significant amounts is
much higher than the mean cloud column density. We know that in clouds dominated by
turbulence, the column density PDF is log-normal, with a typical dispersion around the
mean column density of no more than a factor of two in clouds in which the turbulence
is primarily solenoidal in nature \citep[see e.g.][]{fed10,bl12,abv15}.
Consequently, in turbulence-dominated clouds prior to the
onset of gravitational collapse, there are few or no lines-of-sight associated with bright
CO emission. It follows from this that the mean CO brightness of the cloud is very small,
and hence that $X_{\rm CO}$ is very large, since many of these clouds still contain significant
amounts of H$_{2}$. However, once runaway gravitational collapse begins, the column density
PDF typically develops a pronounced power-law tail at high column densities 
\citep[see e.g.][]{klessen00,kain09,sch15}.
This increases the number of lines-of-sight that have high extinctions and hence significant
CO brightnesses. The result is a dramatic increase in the mean CO brightness of the cloud,
which drives a dramatic decrease in $X_{\rm CO}$. Note that we do not see this behaviour in
high metallicity clouds because in that case, the mean cloud column density already
lies in the range of column densities for which we find bright CO emission. Therefore,
in this case, the cloud is CO-bright prior to the onset of runaway gravitational collapse,
and so $X_{\rm CO}$ does not vary strongly with time.

\subsection{Effect of varying the dust-to-gas ratio}   
In most of our simulations, we assume that the dust-to-gas ratio ${\cal D}$ is
linearly proportional to the metallicity ${\rm Z} / {\rm Z_{\odot}}$. Observationally,
this seems to be a reasonable assumption for metal-rich galaxies like our own
Milky Way \citep{sand13}. However, there is growing evidence that in
metal-poor systems, such as dwarf galaxies, the dust-to-gas ratio becomes a
much steeper function of the gas-phase metallicity \citep[see e.g.][]{gala11,hc12,rr14}.
It is therefore useful to examine what effect this steeper dependence of
${\cal D}$ on ${\rm Z}$ may have on our results for $X_{\rm CI}$ and $X_{\rm CO}$. 

We have therefore performed three additional simulations in which the dust-to-gas
ratio was not simply scaled with  ${\rm Z} / {\rm Z_{\odot}}$. In run Z05-G1-D0375,
we adopted a gas-phase metallicity of ${\rm Z} = 0.5 \: {\rm Z_{\odot}}$ and a
dust-to-gas ratio of ${\cal D} = 0.375 \, {\cal D}_{\odot}$. In run Z02-G1-D01, we 
adopted a gas-phase metallicity ${\rm Z} = 0.2 \: {\rm Z_{\odot}}$ and a dust-to-gas 
ratio ${\cal D} = 0.1 \, {\cal D}_{\odot}$. Finally, in run Z01-G1-D001, we adopted
a gas-phase metallicity ${\rm Z} = 0.1 \: {\rm Z_{\odot}}$ and a dust-to-gas ratio
${\cal D} = 0.01 \, {\cal D}_{\odot}$. The values in the first of these runs were chosen
to be a good match for the gas-phase metallicity and dust-to-gas ratio in the Large
Magellanic Cloud \citep{gordon14}, while the values in the other two runs were chosen
to be consistent with the relationship between metallicity and dust-to-gas ratio inferred
by \citet{rr14}.

\begin{table}
\caption{Comparison of $X_{\rm CI}$ and $X_{\rm CO}$ in runs with low and standard
dust-to-gas ratios \label{tab:d2g}}
\begin{tabular}{lcccc}
\hline
ID &  $X_{\rm CI} (t_{1})$ & $X_{\rm CO}(t_{1})$ & $X_{\rm CI} (t_{\rm SF})$ & $X_{\rm CO}(t_{\rm SF})$ \\
\hline
Z05-G1-D0375 & 4.21 & 2.21 & 7.15 & 1.76 \\
Z05-G1 & 3.85 & 1.79 & 7.05 & 1.69 \\
\hline
Z02-G1-D01     & 18.7 & 35.1 & 16.2 & 3.57 \\
Z02-G1 & 18.2 & 10.9 & 18.6 & 3.13 \\
\hline
Z01-G1-D001   & 9.32 & 3420 & 10.5 & 6.15 \\
Z01-G1 & 54.0 & 183 & 40.9 & 7.75 \\
\hline
\end{tabular}
\\ $X_{\rm CI}$ and $X_{\rm CO}$ are given in units of the Galactic 
CO-to-H$_{2}$ conversion factor $X_{\rm CO, MW}$ (see Eq.~\ref{xcomw}).
Values are shown for $t_{\rm SF}$ and $t_{1} \equiv t_{\rm SF} - 1 \:
{\rm Myr}$.
\end{table}

In Table~\ref{tab:d2g}, we list the values of $X_{\rm CI}$ and $X_{\rm CO}$ in these runs
at $t_{\rm SF}$ and $t_{1} \equiv t_{\rm SF} - 1 \: {\rm Myr}$. For comparison, we also list
the values of $X_{\rm CI}$ and $X_{\rm CO}$ at the same times in runs with the same
metallicity and radiation field strength but with a dust-to-gas ratio scaled linearly with
metallicity.

We see from the table that in run Z05-G1-D0375, the lower dust-to-gas ratio has only
a small effect on the outcome of the simulations. We recover systematically higher
values of $X_{\rm CI}$ and $X_{\rm CO}$ than in run Z05-G1, but only by a small
amount: around 10-20\% at $t_{1}$, and less than 5\% at $t = t_{\rm SF}$.

In run Z02-G1-D01, we also find only minor differences compared to run Z02-G1 if we look
at $X_{\rm CI}$ or at the value of $X_{\rm CO}$ measured at $t_{\rm SF}$. However,
we see that there is a much larger difference in $X_{\rm CO}$ at earlier times: at
$t = t_{1}$, it is roughly three times larger in run Z02-G1-D01 than in run Z02-G1. 
This difference in behaviour is again a consequence of the fact that at $t = t_{1}$,
the CO emission from this cloud is dominated by emission from diffuse gas, while
at $t = t_{\rm SF}$, it is dominated by emission from a single dense core. The
visual extinction of the dense core is sufficiently high that even if we decrease the dust-to-gas
ratio by a factor of two, it still remains well-shielded, allowing CO to resist photodissociation.
Consequently, the change in ${\cal D}$ does not have a strong influence on the value
of $X_{\rm CO}$ that we recover at this point in the evolution of the cloud.  At earlier times,
however, most of the CO emission is produced by gas with a relatively low visual extinction.
The amount of CO that can survive in this gas is much more sensitive to changes in $A_{\rm V}$,
and hence to change in the dust-to-gas ratio, and so it is not surprising that in this regime,
$X_{\rm CO}$ is highly sensitive to the value of ${\cal D}$. 

Finally, in run Z01-G1-D001, we see that the substantial reduction in ${\cal D}$ that is inferred
for very low metallicity systems leads to significant changes in both $X_{\rm CI}$ and $X_{\rm CO}$.
Reducing ${\cal D}$ by a large amount -- a factor of ten in this case -- has two main effects.
First, it significant reduces the amount of H$_{2}$ formed in the cloud, since the H$_{2}$
formation rate is proportional to ${\cal D}$. Second, it reduces the amount of dust shielding,
which decreases the equilibrium abundances of both atomic carbon and CO. In the case of
\ci, the former effect dominates: although the mean \ci~integrated intensity decreases, the
H$_{2}$ column density decreases by a much larger factor, and so $X_{\rm CI}$ decreases.
For CO at early times (i.e.\ prior to runaway gravitational collapse), the reduction in shielding
dominates instead. The CO abundance and CO integrated brightness both decrease by a much
larger factor than the H$_{2}$ column density, and so $X_{\rm CO}$ increases significantly.
Finally, at $t = t _{\rm SF}$, when the total CO emission of the cloud is dominated by the 
contribution from the gravitationally collapsing core, $X_{\rm CO}$ becomes largely independent
of ${\cal D}$, decreasing by only 20\% for a factor of ten decrease in ${\cal D}$.

\subsection{Effect of varying the cosmic ray ionization rate}   
In most of the simulations presented in this paper, we have kept the cosmic ray ionization rate
of atomic hydrogen fixed at its default value of  $\zeta_{\rm H,0} = 10^{-17} \: {\rm s^{-1}}$, even
when varying the strength of the interstellar radiation field. We have done this in order
to allow us to focus solely on the effect of varying one parameter at a time. However, it is likely
that this is not a good approximation for the behaviour of real star-forming systems. The low
energy cosmic rays that make the dominant contribution to $\zeta_{\rm H}$ are believed to be 
produced by Fermi acceleration in supernova remnants. Therefore, the cosmic ray energy density
should scale linearly with the supernova rate, provided that the mean time between supernovae
is shorter than the timescale on which the cosmic rays escape from the galaxy. The supernova
rate in turn depends on the star formation rate, averaged on a 10--20 Myr timescale, and so it is 
reasonable to assume that the cosmic ray energy density, and hence the cosmic ray ionization
rate, should scale linearly with the star formation rate \citep[see e.g.][]{p10,p11}.
The ultraviolet portion of the ISRF that is responsible for the photodissociation of H$_{2}$ and CO
and the photoionization of atomic carbon is dominated by emission from massive stars, and so
should also scale linearly with the star formation rate. Therefore, in a real star-forming system,
it is likely that increases in $G_{0}$ are associated with comparable increases in $\zeta_{\rm H}$.

To investigate the effect that this has on the relative importance of \ci~and CO as tracers of
molecular gas, we have carried out several simulations in which we increased both $G_{0}$
and $\zeta_{\rm H}$ by factors of 10 and 100 compared to our default values. The details
of these runs are listed at the end of Table~\ref{tab:sims}. In five of these runs, star formation begins
after roughly 2--3 Myr, just as in the runs in which we keep $\zeta_{\rm H}$ fixed. In run
Z01-G100-CR100, however, the combination of the elevated ISRF and stronger cosmic
ray flux provide so much heating that star formation is completely suppressed. The gas in
the cloud heats up rapidly, with the mass-weighted mean temperature reaching $T \sim
1000$~K after only 1~Myr, and $T \sim 2400$~K after 2~Myr. This strong heating causes
the cloud to expand, reducing its density and allowing the temperature to climb further, a process
that eventually results in the destruction of the cloud. As a result of this, the C and CO abundances 
remain extremely small throughout the simulation. For this reason, we
do not include this run in our further analysis. 

For the remainder of the runs with elevated cosmic ray ionization rates, we compute $X_{\rm CI}$
and $X_{\rm CO}$ at $t_{\rm SF}$ and $t_{1} \equiv t_{\rm SF} - 1 \: {\rm Myr}$. The results are
shown in Table~\ref{tab:cr}, along with the values from the corresponding runs with fixed $\zeta_{\rm H}$.

\begin{table}
\caption{Comparison of $X_{\rm CI}$ and $X_{\rm CO}$ in runs with different
cosmic ray ionization rates \label{tab:cr}}
\begin{tabular}{lcccc}
\hline
ID &  $X_{\rm CI} (t_{1})$ & $X_{\rm CO}(t_{1})$ & $X_{\rm CI} (t_{\rm SF})$ & $X_{\rm CO}(t_{\rm SF})$ \\
\hline
Z1-G10-CR10 & 2.46 & 2.04 & 4.61 & 2.01 \\
Z1-G10           & 2.95 & 1.47 & 5.85 & 1.58 \\
\hline
Z1-G100-CR100 & 2.87 & 11.7 & 3.18 & 3.02 \\
Z1-G100             & 4.94 & 2.86 & 6.85 & 1.99 \\
\hline
Z03-G10-CR10 & 36.3 & 48.1 & 20.9 & 7.35 \\
Z03-G10           & 54.5 & 35.2 & 33.7 & 6.1 \\
\hline
Z03-G100-CR100 & 85.3 & 2110 & 22.0 & 20.8 \\
Z03-G100             & 326 & 810 & 64.5 & 15.7 \\
\hline
Z01-G10-CR10 & 213 & 30500 & 85.6 & 25.1 \\
Z01-G10             & 463 & 34300 & 132 & 15.0 \\
\hline
\end{tabular}
\\ $X_{\rm CI}$ and $X_{\rm CO}$ are given in units of the Galactic 
CO-to-H$_{2}$ conversion factor $X_{\rm CO, MW}$ (see Eq.~\ref{xcomw}).
Values are shown for $t_{\rm SF}$ and $t_{1} \equiv t_{\rm SF} - 1 \:
{\rm Myr}$.
\end{table}

We see from Table~\ref{tab:cr} that an increase in $\zeta_{\rm H}$ leads to an increase in $X_{\rm CO}$
and a decrease in $X_{\rm CI}$. This behaviour is easy to understand chemically. As we increase
the cosmic ray ionization rate, we increase the amount of He$^{+}$ present in the gas. This in
turn increases the rate at which CO is destroyed by the dissociative charge transfer reaction
\begin{equation}
{\rm CO + He^{+}} \rightarrow {\rm C^{+} + O + He},  \label{cohep}
\end{equation}
which is the dominant destruction mechanism for CO in well-shielded gas. Increasing the
cosmic ray ionization rate therefore reduces the amount of CO present in the cloud, which in
turn reduces the mean CO brightness and increases $X_{\rm CO}$. In well-shielded regions,
the C$^{+}$ produced by this reaction does not persist in the gas for long. Instead, it is converted
to neutral atomic carbon by radiative recombination. The timescale for this process is much shorter
than the timescale on which these carbon atoms are incorporated into new CO molecules, and so
the end result is an increase in the C abundance, and hence an increase in $W_{\rm CI}$ and
decrease in $X_{\rm CI}$. Note also that although our chemical model does not include the
reactions 
\begin{equation}
{\rm C + He^{+}} \rightarrow {\rm C^{+} + He}, \label{chep}
\end{equation}
and
\begin{equation}
{\rm C + H^{+}} \rightarrow {\rm C^{+} + H}, \label{chp}
\end{equation}
we do not expect these reactions to significantly affect the atomic carbon abundance, 
despite the increased H$^{+}$ and He$^{+}$ abundances produced by the higher cosmic
ray ionization rate. The reason for is that, in contrast to reaction~\ref{cohep}, which proceeds
at close to the Langevin rate \citep{ah86}, reactions~\ref{chep} and \ref{chp} occur only very
slowly, since both have rate coefficients that are roughly a million times smaller than that of
reaction~\ref{cohep} \citep{kimura93,stan98}.

Looking in more detail at the values for $X_{\rm CI}$ and $X_{\rm CO}$ listed in Table~\ref{tab:cr},
we see that at $t = t_{\rm SF}$, even relatively large changes in the cosmic ray ionization
rate lead to only comparatively small changes in the conversion factors. $X_{\rm CO}$ 
increases by less than a factor of two, even if we increase $\zeta_{\rm H}$ by a factor of
100. $X_{\rm CI}$ is somewhat more sensitive to the value of $\zeta_{\rm H}$, but even
in this case varies by no more than a factor of three for a factor of 100 change in the
ionization rate. At earlier times in the evolution of the cloud, $X_{\rm CO}$ becomes more
sensitive to $\zeta_{\rm H}$ and $X_{\rm CI}$ becomes slightly less sensitive, but again
the largest differences we see are around a factor of a few. Therefore, the qualitative details
of the results presented in the previous sections are robust against changes in the cosmic
ray ionization rate, even though the quantitative details are not.

\begin{figure}
\includegraphics[width=0.45\textwidth]{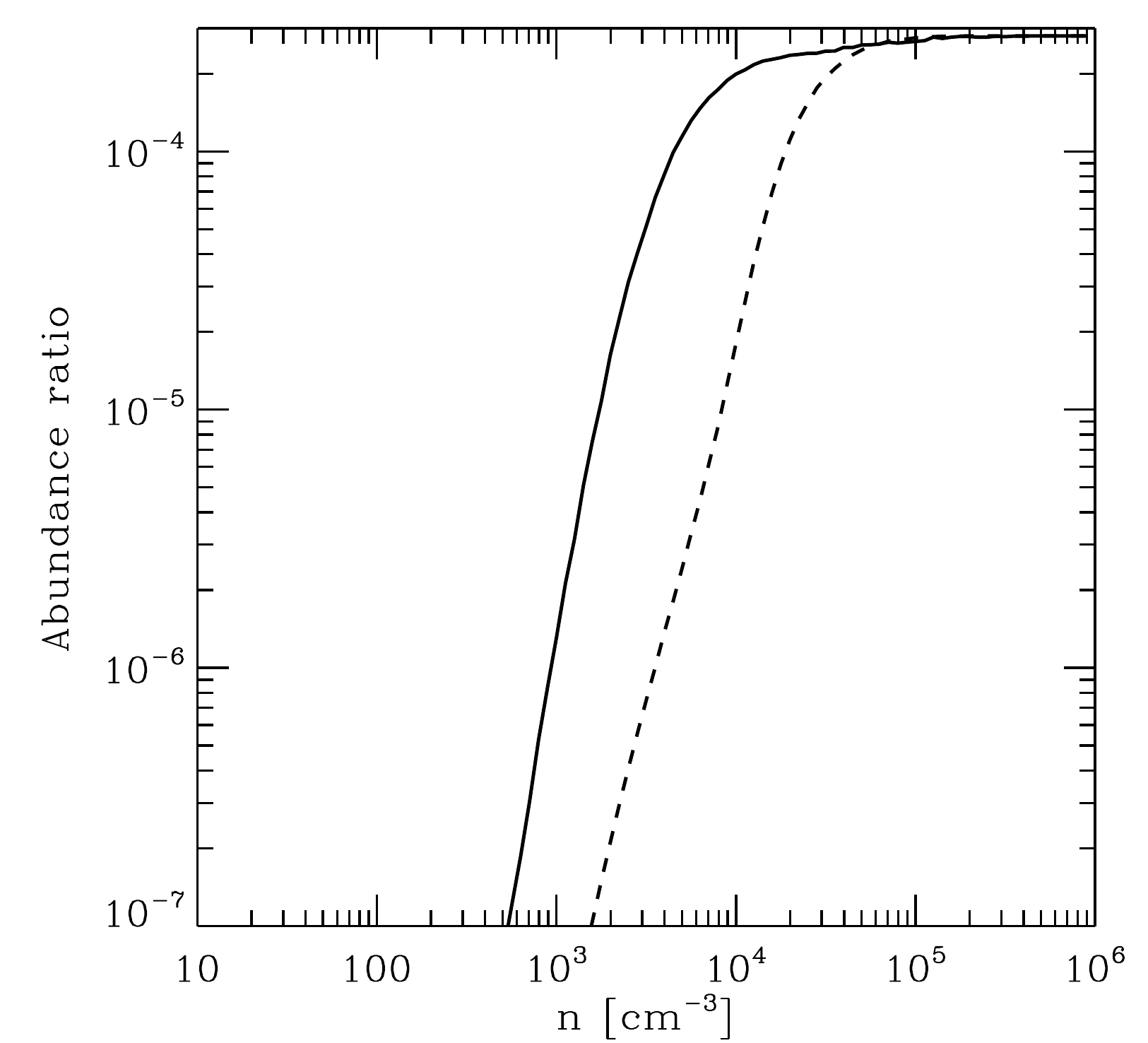}
\caption{Ratio of the mean fractional abundances of CO and H$_{2}$, plotted as a function
of the density, for runs Z1-G100 (solid line) and Z1-G100-CR100 (dashed line) at $t = t_{\rm SF}$. 
Note that in fully molecular gas, our choice for the total carbon abundance implies that
$x_{\rm CO} / x_{\rm H_{2}} = 2.8 \times 10^{-4}$.
\label{fig:COH2}}
\end{figure}   

Finally, it is interesting to compare our results with the recent prediction by \citet{bpv15}
that cosmic ray ionization rates of around $10^{-16}$--$10^{-15} \: {\rm s^{-1}}$ are
sufficient to render molecular gas largely untraceable by CO at typical molecular cloud
densities. As is clear from the results in Table~\ref{tab:cr}, values of $\zeta_{\rm H}$
in this range do increase $X_{\rm CO}$ slightly in evolved clouds, but the clouds are
still bright in CO, which remains a good tracer of molecular gas at metallicities close to
solar. Our results therefore do not confirm the \citet{bpv15} prediction. The discrepancy
between our study and theirs seems to arise from the difference in how we treat the
density structure of the model clouds. \citeauthor{bpv15} largely focus on the effect of 
cosmic rays in uniform density clouds, and find that in this case, at typical GMCs densities of 
100--1000$\, {\rm cm^{-3}}$, CO is strongly suppressed when $\zeta_{\rm H}
\geq 10^{-16} \: {\rm s^{-1}}$ even in the absence of photodissociation.  In our
turbulent clouds, we also find low CO abundances in gas with this range of densities
when $\zeta_{\rm H}$ is large, as illustrated in Figure~\ref{fig:COH2}. In this respect, our model
produces results in good agreement with the \citeauthor{bpv15} results. However, thanks to the influence 
of the turbulent velocity field, the density distribution inside our model clouds is highly inhomogeneous, 
with a significant mass fraction located in regions with a mean density 
substantially higher than the volume-weighted mean density
of the cloud. For example, in run Z1-G100-CR100 at $t = t_{\rm SF}$, almost 10\% of the 
mass of the cloud is located in clumps with $n > 10^{4} \: {\rm cm^{-3}}$, where both
our model and the \citet{bpv15} model agree that the carbon will be almost entirely
locked up in CO. These dense clumps therefore remain CO-bright.
Moreover, since these clumps are distributed
throughout much of the cloud, the mean CO brightness of the cloud as a whole remains
high, with the consequence that $X_{\rm CO}$ remains close to its standard Galactic value.
We can be confident that this is not simply due to some peculiarity of the model
cloud studied here because our results for solar metallicity clouds are in reasonable 
agreement with those presented in the study of \citet{cg15}, which examined a wider range of
cloud masses and densities and used a different random realization of the turbulent
velocity field. Instead, this seems to be a general consequence of the presence of
supersonic turbulence within the clouds.
We note that the fact that the cloud inhomogeneity created by turbulence would allow
more CO to survive was mentioned as a possible caveat by \citet{bpv15}, albeit in the
context of clouds with higher than normal levels of turbulence. Our results show that this
effect is important even in clouds with a `normal' level of turbulence (i.e.\ ones which sit
on the standard size-linewidth relation, as derived by e.g.\ \citealt{larson81} or \citealt{solomon87}).

\subsection{Missing physics}
Our models neglect several physical processes that will be active in real molecular clouds.
The most important of these are the freeze-out of CO onto cold dust grains, the influence
of the magnetic field on the dynamical evolution of the gas, and the effects of star formation
and the associated feedback on the chemical, thermal and dynamical state of the cloud.
In this section, we briefly discuss how these missing processes may affect our results.

\subsubsection{Freeze-out}
In dense gas, the mean time between collisions between a given gas-phase particle
(e.g.\ an H$_{2}$ or CO molecule) and a dust grain can become shorter than the
dynamical time of the gas. In this regime, the freeze-out of gas phase chemicals onto
the surface of dust grains can become important, provided that the average residence
time of these species on the grain surface is also long. The residence time depends
sensitively on the binding energy of the species to the surface, and so H$_{2}$, which
binds only very weakly, does not freeze-out, while CO, which binds much more strongly,
can become significantly depleted from the gas phase \citep[see e.g.][for a good
observational example]{bergin02}.

The simplified chemical model that we use in our simulations does not account for 
the freeze-out of CO, and so the question arises as to whether this leads us to
over-estimate the mean CO integrated intensity, and hence under-estimate $X_{\rm CO}$.
To investigate this, we have attempted to account approximately for the effects of
CO freeze-out in an additional post-processing step. We make two important
simplifying assumptions. First, we assume that locally, the ratio of gas-phase CO
to CO ice has reached equilibrium. This is a reasonable assumption in dense gas,
but means that we will overestimate the degree to which CO is depleted in diffuse
gas, particularly at sub-solar metallicities. Second, we assume that the only processes
that  affect the ratio of gas-phase CO to CO ice are the accretion of CO molecules by dust 
grains, thermal desorption of the CO molecules and cosmic-ray induced desorption 
due to local spot-heating of the grains. Our neglect of photodesorption, which is more 
difficult to model, again means that we will tend to overestimate the extent to which
CO is depleted. 

Having made these assumptions, we can relate the fractional abundance of gas-phase
CO, $x_{\rm CO, gas}$ to the fractional abundance of CO ice, $x_{\rm CO, ice}$ using
the following equation:
\begin{equation}
R_{\rm acc} x_{\rm CO, gas}  = \left(R_{\rm therm} + R_{\rm cr} \right) x_{\rm CO, ice}.
\end{equation}
Here, $R_{\rm acc}$ is the rate per CO molecule at which CO is accreted by dust grains,
$R_{\rm therm}$ is the rate per CO molecule at which CO evaporates from the grain
surfaces and $R_{\rm cr}$ is the rate per CO molecule at which CO is desorbed by
cosmic rays. Following \citet{hollen09}, we have
\begin{equation}
R_{\rm acc} = 10^{-17} T^{1/2} {\cal D} n  \: \: {\rm s^{-1} \: molecule^{-1}},  \label{racc}
\end{equation}
where $n$ is the number density of hydrogen nuclei, 
\begin{equation}
R_{\rm therm} = 1.04 \times 10^{12} \exp \left(\frac{-960}{T_{\rm d}}\right) \: \: {\rm s^{-1} \: molecule^{-1}},
\end{equation}
where $T_{\rm d}$ is the dust temperature, and
\begin{equation}
R_{\rm cr} = 6.0 \times 10^{-13} \zeta_{\rm H, 17} \: \: {\rm s^{-1} \: molecule^{-1}},  \label{rcr}
\end{equation}
where $\zeta_{\rm H, 17} = \zeta_{\rm H} / 10^{-17} \: {\rm s^{-1}}$.
The expression here for the thermal desorption rate assumes a CO adsorption energy of
$E_{\rm D} / k = 960$~K, following \citet{aikawa96}. The cosmic ray desorption rate is
based on the calculations of \citet{hc06}. 

Armed with the depletion and desorption rates, we can write the ratio of gas-phase
CO to CO ice as
\begin{equation}
\frac{x_{\rm CO, gas}}{x_{\rm CO, ice}} = \frac{R_{\rm therm} + R_{\rm cr}}{R_{\rm acc}}.
\end{equation}
For $T_{\rm d} \ll 17$~K, cosmic-ray desorption dominates over thermal desorption and 
this reduces to
\begin{equation}
\frac{x_{\rm CO, gas}}{x_{\rm CO, ice}} = \frac{R_{\rm cr}}{R_{\rm acc}} \simeq 6 \times 10^{4} T^{-1/2} \frac{\zeta_{\rm H, 17}}{{\cal D} n}.
\end{equation}
In dense gas with $T = 15$~K, ${\cal D} = 1$ and $\zeta_{\rm H, 17} = 1$, we therefore expect that roughly half
of the CO will be depleted once the gas density reaches $n \sim 15000 \: {\rm cm^{-3}}$.
Observationally, the density at which $x_{\rm CO, gas} \sim x_{\rm CO, ice}$ is found to be a factor of a few
larger than this \citep{lippok13}, demonstrating that, as expected, our simplified model overestimates the
effectiveness of CO freeze-out.

To investigate the effect that freeze-out has on our predicted CO integrated intensities, we use the simplified
model outlined above as part of an additional post-processing step. Before interpolating the CO densities from
our SPH particles to the Cartesian grid that we use for our radiative transfer calculations, we first rescale them
by a factor
\begin{equation}
F_{\rm gas} = \frac{x_{\rm CO, gas}}{x_{\rm CO, gas} + x_{\rm CO, ice}} = \frac{R_{\rm therm} + R_{\rm cr}}{R_{\rm therm} + R_{\rm cr} + R_{\rm acc}}
\end{equation}
to account for the effects of freeze-out, using the expressions for $R_{\rm acc}$, $R_{\rm therm}$ and
$R_{\rm cr}$ given in Equations~\ref{racc}--\ref{rcr} above.
We then proceed as before, computing the CO emission from the cloud using {\sc radmc-3d}, and 
then using the mean values of the CO integrated intensity and H$_{2}$ column density to compute
$X_{\rm CO}$. 

\begin{table}
\caption{Comparison of $X_{\rm CO}$ computed with and without our approximate freeze-out model
\label{tab:freeze}}
\begin{tabular}{lcc}
\hline
ID &  \multicolumn{2}{c}{$X_{\rm CO}$} \\
& Without freeze-out & With freeze-out \\
\hline
Z1-G1     & 1.33 & 1.36 \\
Z02-G1   & 3.13 & 3.32 \\
Z003-G1 & 21.9 & 28.5 \\
\hline
\end{tabular}
\\ $X_{\rm CO}$ is given in units of the Galactic 
CO-to-H$_{2}$ conversion factor $X_{\rm CO, MW}$ (see Eq.~\ref{xcomw})
and is computed for $t = t_{\rm SF}$.
\end{table}

We have carried out this analysis for three of our cloud models, Z1-G1, Z02-G1 and Z003-G1, chosen 
to span the full range of metallicities examined in this study. We list the resulting values of $X_{\rm CO}$
in Table~\ref{tab:freeze}, along with the values that we obtain in the absence of freeze-out. We see that
at solar metallicity, freeze-out has a negligible effect on the predicted value of $X_{\rm CO}$. This is easy
to understand, as by the time we reach the densities at which CO freeze-out becomes significant, the gas
is highly optically thick in the $^{12}$CO $1 \rightarrow 0$ line. Because the line is optically thick, removing
even a large fraction of the CO has little effect on the total emission (provided that $\tau$ remains greater
than one), and hence little effect on $X_{\rm CO}$. 

As we reduce the metallicity, we find that freeze-out starts to have a greater effect on $X_{\rm CO}$.
Again, this is easy to understand: at lower ${\rm Z}$, we have much less CO and hence the optical
depth of the line is much smaller, making it more sensitive to the actual CO abundance. On the other
hand, reducing the dust-to-gas ratio, which is assumed in these models to scale with Z, reduces the
freeze-out rate, since there is less dust for the CO molecules to freeze out onto. Consequently, in the
lower metallicity models, the density at which $x_{\rm CO, gas} \sim x_{\rm CO, ice}$ is much higher
than in run Z1-G1. As a result, freeze-out remains relatively unimportant in these models despite
the lower optical depths, increasing $X_{\rm CO}$ by at most 30\%. 

Finally, what about the other simulations studied in this paper? In the other runs with $G_{0} = 1$,
we would expect to recover results intermediate between the Z1-G1 and Z003-G1 results, with the
precise behaviour depending on the metallicity. In the runs with reduced dust-to-gas ratio, the 
effects of freeze-out will be smaller, since $R_{\rm acc} \propto {\cal D}$. In the runs with 
$G_{0} = 10$ or 100, we would again expect the effects of freeze-out to be smaller, since the 
dust will be warmer, making $R_{\rm therm}$ larger. Finally, in the runs with elevated cosmic ray
ionization rates, $R_{\rm cr}$ will be larger, and so once again we will have less freeze-out. 
Therefore, it seems safe to conclude that the freeze-out of CO onto dust will not strongly affect
the results we recover for the behaviour of $X_{\rm CO}$ in any of our cloud models.

\subsubsection{Magnetic fields}
Real molecular clouds are known to be magnetized \citep[see e.g.][]{crutcher12}, but we carry out our simulations
using a purely hydrodynamical approach, in view of the technical difficulties involved in
modeling magnetized gas with smoothed particle hydrodynamics \citep[see the lengthy discussion in][]{price12}. 
The question therefore arises as to how much this is likely to affect our results. 

We know from previous simulations of turbulent molecular clouds \citep[e.g.][]{pn11,molina12} that prior to the
onset of runaway gravitational collapse, the width of the volume density and column density
PDFs are sensitive to the strength of the magnetic field. Increasing the field strength provides
additional resistance against compression or rarefaction and leads to a narrower volume 
density PDF. Projection of this into two dimensions then leads in turn to a narrower column
density PDF. The extent to which this will affect $X_{\rm CI}$ and $X_{\rm CO}$ will then depend
on the size of the characteristic column density (or visual extinction) at which we find the
transition from C$^{+}$ to C or from C to CO. If this column density is comparable to or smaller 
than the mean column density of the cloud, then a significant fraction of the gas will be \ci~or
CO rich, regardless of whether we consider a magnetized or unmagnetized cloud. In this
case, we would expect the values of $X_{\rm CI}$ and $X_{\rm CO}$ to be insensitive to
the magnetic field strength. Similarly, 
if the required column density is so large that it is reached only in regions that go into
runaway gravitational collapse (as in e.g.\ our low metallicity runs), then the amount of
C and CO that form in the cloud will be insensitive to the width of the turbulence-dominated
density PDF, implying that $X_{\rm CI}$ and $X_{\rm CO}$ will also not vary much.
The only case in which the difference in the width of the PDF is likely to have a large effect
is if the characteristic column density is such that it is 1--2 standard deviations above the
mean in the unmagnetized cloud, but multiple standard deviations above the mean in the
magnetized cloud, owing to the change in the size of the standard deviation. In this case,
we would go from predicting that the cloud is \ci~and CO-rich in the unmagnetized 
case to predicting that it is \ci~and CO-poor in the magnetized case. However, such
a situation will rarely be encountered in practice, as magnetic fields of the level observed
in most GMCs do not make a large difference to the width of the column density PDF.
We therefore
expect that accounting for this effect may lead to minor changes in our results at 
intermediate metallicities, but that it will not greatly change our main conclusions regarding the
utility of \ci~and CO as tracers of molecular gas.

The other potentially important effect of the magnetic field will be to suppress the 
gravitational collapse of dense regions, if they are magnetically sub-critical. This
can have a large effect on $X_{\rm CO}$ in low metallicity clouds: if 
gravitational collapse is suppressed, the gas will never become CO-bright, and
so $X_{\rm CO}$ will remain large. Clouds that are so magnetically dominated
that they remain starless for long periods will therefore also remain invisible in CO 
for long periods. In this circumstance, \ci~will clearly be a much better tracer of the
clouds. However, it should also be noted that the importance of this effect is
unclear. Measurements of the magnetic field strength in dense prestellar cores in
local molecular clouds suggest that these cores are generally magnetically 
supercritical, and hence that gravitational collapse is not strongly suppressed
\citep{tc08,crutcher12}. Whether the same holds true for dense clouds in much lower
metallicity environments is unknown.

\subsubsection{Star formation and feedback}
Our simulations do not account for the effects of stellar feedback. Because of this, we cannot
explore how $X_{\rm CI}$ and $X_{\rm CO}$ evolve over the entire lifetime of one of our
simulated clouds, as without feedback, there is nothing to stop the whole of the gas 
eventually being converted to stars. We have therefore focussed primarily on their
evolution prior to the onset of star formation (i.e.\ at $t < t_{\rm SF}$), when stellar
feedback is, by definition, completely absent. 

However, in Section~\ref{sec:time_evol} we did investigate how $X_{\rm CO}$ and $X_{\rm CI}$ evolved
in the interval between $t = t_{\rm SF}$ and $t = t_{\rm SF} + 0.5 \: {\rm Myr}$ in a few
of our simulations, and it is therefore important to note that our results for this regime 
may be affected by our neglect of feedback. The importance of feedback during this period 
will depend on the star formation rate of the clouds and how much of the dense, CO-bright
gas is actively forming stars. For example, in run Z1-G1, the star formation efficiency of the
cloud at $t = t_{\rm SF} + 0.5 \: {\rm Myr}$ is only around 0.3\%, and much of the CO is located
in gas which is too diffuse to be actively star-forming. In this cloud, we would not expect the
evolution of $X_{\rm CO}$ and $X_{\rm CI}$ at $t > t_{\rm SF}$ to have yet been strongly
affected by stellar feedback. On the other hand, in run Z003-G100, the star formation 
efficiency of the cloud at $t = t_{\rm SF} + 0.5 \: {\rm Myr}$ is around 7.8\%, more than an
order of magnitude larger than in the solar metallicity case. This difference in star formation
efficiency is a consequence of the different thermal structure of the two clouds. The gas in
the low metallicity, strongly irradiated cloud is significantly warmer than the gas in the solar
metallicity cloud. This means that it takes longer to form a dense core which is sufficiently
massive to be self-gravitating (see Table~\ref{tab:tsf}), but also that once such a core has formed,
it is able to form a lot of stars relatively rapidly. Because of the high instantaneous star 
formation rate in this cloud, and the fact that all of the CO is located in a single dense core,
it is plausible that if we were to include the effects of stellar feedback, much of the CO would
be destroyed. In this case, the period during which the cloud is CO-bright would be even
more short-lived that we find in our simulations. However, confirmation of this requires
simulations that do account for the effects of stellar feedback, which are out of the scope
of our present study.

\section{Conclusions}
\label{conc}
The most important point to take away from this study is that the usefulness of \ci~emission as a tracer
of H$_{2}$ mass in low metallicity clouds depends strongly on the evolutionary state of the
clouds. At early times, prior to the onset of gravitational collapse in the cloud, \ci~is a far 
better tracer of H$_{2}$ than CO. The \ci-to-H$_{2}$ conversion factor, $X_{\rm CI}$, is
not constant -- it varies with the metallicity roughly as $X_{\rm CI} \propto {\rm Z}^{-1.5}$--${\rm Z}^{-2}$
for metallicities in the range $0.1 < {\rm Z / Z_{\odot}} < 1.0$ -- but it is far less sensitive
to changes in the metallicity than CO, which is an extremely ineffective tracer of molecular
mass in starless clouds at low metallicity. 

Shortly before the onset of star formation, however, CO becomes a much better tracer of
the molecular mass. At $t = t_{\rm SF}$, both $X_{\rm CO}$ and $X_{\rm CI}$ scale 
approximately with metallicity as $X \propto {\rm Z^{-1}}$, but $X_{\rm CO}$ is consistently
smaller than $X_{\rm CI}$, by a factor ranging from a few to an order of magnitude. Since
it is much easier to observe the $^{12}$CO $J = 1 \rightarrow 0$ line in terrestrial conditions
than the \ci~609$\,\mu$m line, the conclusion we draw from this is that at redshifts close to
zero, CO should be preferred to \ci~as a tracer of the H$_{2}$ mass of star-forming clouds in low 
metallicity systems.\footnote{Arguably, the [C$\,${\sc ii}] 158$\,\mu$m fine structure line is an even
better tracer of H$_{2}$ in these systems \citep[see e.g.][]{cormier15}, but this topic lies outside of the 
scope of this paper.} At higher redshifts, however, the \ci~609$\,\mu$m line becomes much easier
to observe from the ground, and in this case \ci~can become competitive with CO as a tracer
of molecular gas in metal-poor star-forming systems \citep{ptv04}.

The difference in the behaviour of starless and star-forming clouds can be understood as
a consequence of the physical conditions required to form CO in low metallicity clouds.
Substantial amounts of CO form only in gas with $A_{\rm V} \sim 1$ or above. In
low metallicity clouds, this gas is found only in dense, gravitationally collapsing cores.
At early times in the evolution of the cloud, its structure is dominated by turbulence rather
than gravity, and these dense cores are not present. Consequently, the mean CO brightness 
of the cloud at this time is extremely small. At later times, however, dense cores form due
to gravitational instability, leading to a dramatic increase in the CO brightness of the cloud.
On the other hand, the \ci~abundance is far less sensitive to the visual extinction, and so
the mean \ci~brightness of the cloud varies much less as the cloud evolves.

We have also examined how our results change if we relax our assumption that the dus-to-gas
ratio ${\cal D}$ scales linearly with the gas-phase metallicity, and instead allow it to drop more
rapidly with decreasing ${\rm Z}$, as suggested by the available observational data. We find
that in practice, this strengthens our results regarding the relative usefulness of CO and 
\ci~as molecular gas tracers. Significantly decreasing ${\cal D}$ leads to a substantial 
decrease in the CO content of the clouds at early times, and hence a substantial increase in
$X_{\rm CO}$. On other other hand, it has little influence on the value of $X_{\rm CO}$ that
we recover once star formation begins. At the same time, reducing ${\cal D}$ also leads to
a decrease in the amount of H$_{2}$ formed in the cloud, which acts to decrease $X_{\rm CO}$,
although this effect only seems to be important if the dust-to-gas ratio is very low.

Finally, we have examined the consequences of allowing the cosmic ray ionization rate to
increase at the same time as the strength of the ISRF, as we would expect to occur if both
directly track the star formation rate \citep[see e.g.][]{cg15}. We find that qualitatively, we
recover similar results, but some of the quantitative details change. Notably, increasing the
cosmic ray ionization rate by a factor of 10 or 100 has only a minor effect on the values of
$X_{\rm CO}$ we recover for our star-forming clouds (contrary to the prediction of \citealt{bpv15}), 
but can lead to a factor of two to three decrease in $X_{\rm CI}$, improving our prospects for
using \ci~emission to trace H$_{2}$ in these systems.

\section*{Acknowledgements}
The authors thank the anonymous referee for a constructive report that helped them to
improve the paper. They also
acknowledge useful conversations with D.~Cormier, S.~Madden, and 
S.~Malhotra on the topic of \ci~in metal-poor galaxies. SCOG acknowledges financial support 
from the Deutsche Forschungsgemeinschaft  via SFB 881, ``The Milky Way System'' 
(sub-projects B1, B2 and B8) and SPP 1573, ``Physics of the Interstellar Medium'' 
(grant number GL 668/2-1).


\begin{thebibliography}{}

\bibitem[Abreu-Vicente et~al.(2015)]{abv15}
Abreu-Vicente, J., Kainulainen, J., Stutz, A., Henning, Th., \& Beuther, H.\ 2015, A\&A, 581, A74

\bibitem[Aikawa et~al.(1996)]{aikawa96}
Aikawa, Y., Miyama, S.~M., Nakano, T., \& Umebayashi, T.\ 1996, ApJ, 467, 684

\bibitem[Anicich \& Huntress(1986)]{ah86}
Anicich, V.~G., \& Huntress, W.~T.\ 1986, ApJS, 62, 553

\bibitem[Bate, Bonnell \& Price(1995)]{bbp95}
Bate, M.~R., Bonnell, I.~A., \& Price, N.~M.\ 1995, MNRAS, 277, 362

\bibitem[Benz(1990)]{benz90}
Benz, W., 1990, in `Proceedings of the NATO Advanced Research Workshop on The
Numerical Modelling of Nonlinear Stellar Pulsations Problems and Prospects',
ed.\ J.~R.~Buchler, (Dordrecht: Kluwer), 269

\bibitem[Bergin et~al.(2002)]{bergin02}
Bergin, E.~A., Alves, J., Huard, T., \& Lada, C.~J.\ 2002, ApJ, 570, L101

\bibitem[Beuther et~al.(2014)]{beuther14}
Beuther, H., Ragan, S.~E., Ossenkopf, V., Glover, S., Henning, Th., Linz, H., Nielbock, M., Krause, O.,
Stutzki, J., Schilke, P., \& G\"usten, R.\ 2014, A\&A, 571, 53

\bibitem[Bisbas et~al.(2012)]{bisbas12}
Bisbas, T.~G., Bell, T.~A., Viti, S., Yates, J., \& Barlow, M.~J.\ 2012, MNRAS, 427, 2100

\bibitem[Bisbas, Papadopoulos, \& Viti(2015)]{bpv15}
Bisbas, T.~G., Papadopoulos, P.~P., \& Viti, S.\ 2015, ApJ, 803, 37

\bibitem[Bolatto et~al.(2011)]{bolatto11}
Bolatto, A., et~al., 2011, ApJ, 741, 12

\bibitem[Bolatto, Wolfire \& Leroy(2013)]{bwl13}
Bolatto, A.~D., Wolfire, M., Leroy, A.~K.\ 2013, ARA\&A, 51, 207 

\bibitem[Burkhart \& Lazarian(2012)]{bl12}
Burkhart, B., \& Lazarian, A.\ 2012, ApJ, 755, L19

\bibitem[Clark, Glover \& Klessen(2012)]{cgk12}
Clark, P.~C., Glover, S.~C.~O., \& Klessen, R.~S.\ 2012, MNRAS, 420, 745

\bibitem[Clark \& Glover(2015)]{cg15}
Clark, P.~C., \& Glover, S.~C.~O., 2015, MNRAS, 452, 2057

\bibitem[Crutcher(2012)]{crutcher12}
Crutcher, R.~M.\ 2012 ARA\&A, 50, 29

\bibitem[Cormier et~al.(2015)]{cormier15}
Cormier, D., et~al., 2015, A\&A, 578, A53

\bibitem[Draine(1978)]{dr78}
Draine, B.~T. 1978, ApJS,  36, 595

\bibitem[Elmegreen et al.(2013)]{elm13}
Elmegreen, B.~G., Rubio, M., Hunter, D.~A., Verdugo, C., Brinks, E., \& Schruba, A.\
2013, Nature, 495, 487

\bibitem[Federrath et~al.(2010)]{fed10}
Federrath, C., Roman-Duval, J., Klessen, R.~S., Schmidt, W., \& {Mac Low}, M.-M.\ 2010, A\&A, 512, A81

\bibitem[\protect\citeauthoryear{Frerking et al.}{1989}]{frerking89}
Frerking, M.~A., Keene, J., Blake, G.~A., Phillips, T.~G., 1989, ApJ, 344, 311

\bibitem[Galametz et~al.(2011)]{gala11}
Galametz, M., Madden, S.~C., Galliano, F., Hony, S., Bendo, G.~J., \& Sauvage, M.\
2011, A\&A, 532, 56 

\bibitem[Genzel et~al.(1988)]{genz88}
Genzel, R., Harris, A.~I., Stutzki, J., \& Jaffe, D.~T.\ 1988, ApJ, 332, 1049

\bibitem[Glover \& Clark(2012a)]{gc12a}
Glover, S.~C.~O., \& Clark, P.~C.\ 2012a, MNRAS, 421, 9

\bibitem[Glover \& Clark(2012b)]{gc12b}
Glover, S.~C.~O., \& Clark, P.~C.\ 2012b, MNRAS, 426, 377

\bibitem[Glover \& Clark(2012c)]{gc12c}
Glover, S.~C.~O., \& Clark, P.~C.\ 2012c, MNRAS, 421, 116

\bibitem[Glover \& {Mac Low}(2007a)]{gm07a}
Glover, S.~C.~O., \& {Mac Low}, M.-M.\ 2007a, ApJS, 169, 239

\bibitem[Glover \& {Mac Low}(2007b)]{gm07b}
Glover, S.~C.~O., \& {Mac Low}, M.-M.\ 2007b, ApJ, 659, 1317

\bibitem[Glover \& {Mac Low}(2011)]{gm11}
Glover, S.~C.~O., \& {Mac Low}, M.-M.\ 2011, MNRAS, 412, 337

\bibitem[Glover et~al.(2015)]{glo15}
Glover, S.~C.~O., Micic, M., Molina, F., \& Clark, P.~C.\ 2015, MNRAS, 448, 1607

\bibitem[Glover et al.(2010)]{glo10}
Glover, S.~C.~O., Federrath, C., {Mac Low}, M.-M., \& Klessen, R.~S.\ 2010, MNRAS, 404, 2

\bibitem[Gordon et~al.(2014)]{gordon14}
Gordon, K.~D., et~al.\ 2014, ApJ, 797, 85

\bibitem[Gratier et~al.(2010)]{gratier10}
Gratier, P., et~al., 2010, A\&A, 512, 68

\bibitem[Herbst \& Cuppen(2006)]{hc06}
Herbst, E., \& Cuppen, H.~M.\ 2006, Proc.\ Natl.\ Acad.\ Sci., 103, 12257

\bibitem[Herrera-Camus et~al.(2012)]{hc12}
Herrera-Camus, R., et~al., 2012, ApJ, 752, 112

\bibitem[Hollenbach et al.(2009)]{hollen09}
Hollenbach, D., Kaufman, M.~J., Bergin, E.~A., \& Melnick, G.~J.\ 2009, ApJ, 690, 1497

\bibitem[Israel(1997)]{israel97}
Israel, F.~P.\ 1997, A\&A, 328, 471

\bibitem[Jappsen et~al.(2005)]{jap05}
Jappsen, A.-K., Klessen, R.~S., Larson, R.~B., Li, Y., {Mac Low}, M.-M.\ 2005,
A\&A, 435, 611

\bibitem[Kainulainen et~al.(2009)]{kain09}
Kainulainen, J., Beuther, H., Henning, T., \& Plume, R.\ 2009, A\&A, 508, L35

\bibitem[Kimura et~al.(1993)]{kimura93}
Kimura, M., Dalgarno, A., Chantranupong, L., Li, Y., Hirsch, G., \& Buenker, R.~J.\ 1993, ApJ, 417, 812

\bibitem[Klessen(2000)]{klessen00}
Klessen, R.~S., 2000, ApJ, 535, 869

\bibitem[Kramer et~al.(2004)]{kram04}
Kramer, C., Jakob, H., Mookerjea, B., Schneider, N., Br\"ull, M., \& Stutzki, J.\ 2004, A\&A, 424, 887

\bibitem[Kramer et~al.(2008)]{kra08}
Kramer, C., et~al., 2008, A\&A, 477, 547

\bibitem[\protect\citeauthoryear{Langer}{1976}]{langer76}
Langer, W., 1976, ApJ, 206, 699

\bibitem[Larson(1981)]{larson81}
Larson, R.~B., 1981, MNRAS, 194, 809

\bibitem[Lee et~al.(2014)]{lee14}
Lee, M.-Y., Stanimirovi\'c, S., Wolfire, M.~G., Shetty, R., Glover, S.~C.~O., Molina, F.~Z., \& Klessen, R.~S.\ 
2014, ApJ, 784, 80

\bibitem[Leroy et al.(2007)]{leroy07}
Leroy, A., Bolatto, A., Stanimirovic, S., Mizuno, N., Israel, F., \& Bot, C.\ 2007, ApJ, 658, 1027

\bibitem[Leroy et al.(2011)]{leroy11}
Leroy, A., et al., 2011, ApJ, 737, 12

\bibitem[Lippok et~al.(2013)]{lippok13}
Lippok, N., et~al.\ 2013, A\&A, 560, A41

\bibitem[\protect\citeauthoryear{Little et al.}{1994}]{little94}
Little, L.~T., Gibb, A.~G., Heaton, B.~D., Ellison, B.~N., Claude, S.~M.~X., 1994, MNRAS, 271, 649

\bibitem[Madden et~al.(1997)]{madden97}
Madden, S.~C., Poglitsch, A., Geis, N., Stacey, G.~J., \& Townes, C.~H., 1997, ApJ, 483, 200

\bibitem[Mathis, Mezger \& Panagia(1983)]{mmp83} 
Mathis, J.~S., Mezger, P.~G., \& Panagia, N. 1983,  A\&A, 128, 212

\bibitem[Molina et al.(2012)]{molina12}
Molina, F.~Z., Glover, S.~C.~O., Federrath, C., \& Klessen, R.~S.\ 2012, MNRAS, 423, 2680

\bibitem[Narayanan et~al.(2011)]{nara11}
Narayanan, D., Krumholz, M., Ostriker, E.~C., \& Hernquist, L.\ 2011, MNRAS, 418, 664

\bibitem[Narayanan \& Hopkins(2013)]{nh13}
Narayanan, D., Hopkins, P.~F., 2013, MNRAS,  433, 1223

\bibitem[Nelson \& Langer(1999)]{nl99}
Nelson, R.~P., \& Langer, W.~D. 1999, ApJ, 524, 923

\bibitem[Offner et~al.(2013)]{offner13}
Offner, S.~S.~R., Bisbas, T.~G., Viti, S., \& Bell, T.~A.\ 2013, ApJ, 770, 49

\bibitem[Offner et~al.(2014)]{offner14}
Offner, S.~S.~R., Bisbas, T.~G., Bell, T.~A., \& Viti, S.\ 2014, MNRAS, 440, L81

\bibitem[Padoan \& Nordlund(2011)]{pn11}
Padoan, P., \& Nordlund, \AA, 2011, ApJ, 730, 40

\bibitem[Papadopoulos(2010)]{p10}
Papadopoulos, P.~P., 2010, ApJ, 720, 226

\bibitem[Papadopoulos, Thi \& Viti(2004)]{ptv04}
Papadopoulos, P.~P., Thi, W.-F., \& Viti, S.\ 2004, MNRAS, 351, 147

\bibitem[Papadopoulos et~al.(2011)]{p11}
Papadopoulos, P.~P., Thi, W.-F., Miniati, F., \& Viti, S.\ 2011, MNRAS, 414, 1705

\bibitem[Pineda, Caselli \& Goodman(2008)]{pineda08}
Pineda, J.~E., Caselli, P., \& Goodman, A.~A.\ 2008, ApJ, 679, 481

\bibitem[Plume, Jaffe \& Keene(1994)]{pjk94}
Plume, R., Jaffe, D.~T., \& Keene, J.\ 1994, ApJ, 425, L49

\bibitem[Price(2012)]{price12}
Price, D.~J.\ 2012, J.\ Comp.\ Phys., 231, 759

\bibitem[R\'emy-Ruyer et~al.(2014)]{rr14}
R\'emy-Ruyer, A., et~al.\ 2014, A\&A, 563, A31

\bibitem[Roman-Duval et~al.(2010)]{RomanDuval2010}
Roman-Duval, J., Jackson, J.~M., Heyer, M., Rathborne, J., \& Simon, R.\ 2010, ApJ, 723, 492

\bibitem[Sandstrom et~al.(2013)]{sand13}
Sandstrom, K.~M., et~al.\ 2013, ApJ, 777, 5

\bibitem[\protect\citeauthoryear{Schilke et al.}{1995}]{sch95}
Schilke, P., Keene, J., Le Bourlot, J., Pineau des Forets, G., Roueff, E., 1995, A\&A, 294, L17

\bibitem[Schneider et~al.(2003)]{sch03}
Schneider, N., Simon, R., Kramer, C., Kraemer, K., Stutzki, J., \& Mookerjea, B.\ 2003, A\&A, 406, 915

\bibitem[Schneider et~al.(2015)]{sch15}
Schneider, N., Csengeri, T., Klessen, R.~S., Tremblin, P., Ossenkopf, V., Peretto, N., Simon, R., Bontemps, S., \& Federrath, C.\
2015, A\&A, 578, A29

\bibitem[Sembach et~al.(2000)]{sem00}
Sembach, K.~R., Howk, J.~C., Ryans, R.~S.~I., \& Keenan, F.~P. 2000, ApJ,  528, 310

\bibitem[Shetty et~al.(2011a)]{shetty11a}
Shetty, R., Glover, S.~C., Dullemond, C., \& Klessen, R.~S. 2011a, MNRAS, 412, 1686

\bibitem[Shetty et~al.(2011b)]{shetty11b}
Shetty, R., Glover, S.~C.~O., Dullemond, C.~P., Ostriker, E.~C., Harris, A.~I.,
\& Klessen, R.~S., 2011b, MNRAS, 415, 3253

\bibitem[Shi et~al.(2015)]{shi15}
Shi, Y., Wang, J., Zhang, Z.-Y., Gao, Y., Armus, L., Helou, G., Gu, Q., \& Stierwalt, S.\
2015, ApJ, 804, L11

\bibitem[Smith et~al.(2012)]{smith12}
Smith, M.~W.~L., et~al., 2012, ApJ, 756, 40

\bibitem[Solomon et~al.(1987)]{solomon87}
Solomon, P.~M., Rivolo, A.~R., Barrett, J., \& Yahil, A.\ 1987, ApJ, 319, 730

\bibitem[Springel(2005)]{springel05}
Springel, V.\ 2005, MNRAS, 364, 1105

\bibitem[Stancil et~al.(1998)]{stan98}
Stancil, P.~C., Havener, C.~C., Krst\'ic, P.~S., Schultz, D.~R., Kimura, M., Gu, J.-P., Hirsch, G., Buenker, R.~J., \&
Zygelman, B.\ 1998, ApJ, 502, 1006

\bibitem[Stutzki et~al.(1988)]{stut88}
Stutzki, J., Stacey, G.~J., Genzel, R., Harris, A.~I., Jaffe, D.~T., \& Lugten, J.~B.\ 1988, ApJ, 332, 379

\bibitem[Troland \& Crutcher(2008)]{tc08}
Troland, T.~H., \& Crutcher, R.~M.\ 2008, ApJ, 680, 457

\bibitem[Woodall et~al.(2007)]{umist07}
Woodall, J., Ag\'undez, M., Markwick-Kemper, A.~J., \& Millar, T.~J.\ 2007, A\&A, 466, 1197

\end{thebibliography}
\end{document}